\documentclass[12pt]{iopart}
\usepackage{mathrsfs}
\usepackage{graphicx}
\usepackage{bm}
\expandafter\let\csname equation*\endcsname\relax
\expandafter\let\csname endequation*\endcsname\relax
\usepackage{amsmath}
\usepackage{epstopdf}
\usepackage{subcaption}
\usepackage{xcolor}
\usepackage{CJK}

\usepackage{iopams}

\begin{document}

\title{Anisotropic odd elasticity with Hamiltonian curl forces}
\begin{CJK*}{UTF8}{gbsn}
\author{Yi-Heng Zhang (张一恒) and Zhenwei Yao$^{\ast}$ (姚振威)}
\address{School of Physics and Astronomy, and Institute of Natural
Sciences, Shanghai Jiao Tong University, Shanghai 200240, China}
\end{CJK*}
\ead{zyao@sjtu.edu.cn}

\begin{abstract}
A host of elastic systems consisting of active components exhibit
path-dependent elastic behaviors not found in classical elasticity, which is
known as odd elasticity.  Odd elasticity is characterized by antisymmetric
(odd) elastic modulus tensor. Here, from the perspective of geometry, we
construct the Hamiltonian formalism to show the origin of the antisymmetry
of the elastic modulus that is intrinsically anisotropic. Furthermore, both non-conservative stress and the associated
nonlinear constitutive relation naturally arise. This work also opens the
promising possibility of exploring the physics of odd elasticity in
dynamical regime by Hamiltonian formalism.
\end{abstract}

%
\vspace{2pc}

\section{Introduction}
In classical elasticity, the input work to deform the elastic body depends only
on its initial and final states. Such path-independence is characterized by the
elastic potential that yields conservative
stress~\cite{Truesdell1952,Landau1960}. Recently, it has been reported
that the work involved in deformations is dependent on the path in a class of
elastic systems consisting of active components, such as in robotic
metamaterials~\cite{Chen2021,Brandenbourger2021}. A common feature in these
active systems is that nonzero work could be extracted in a cycle of
deformation. The dependence of the work on the deformation path could be
attributed to an additional antisymmetric (odd) part in the elastic modulus
tensor, which is known as odd
elasticity~\cite{Salbreux2017,Scheibner2020,fruchart2022}.  The broken major
symmetry of odd-elastic modulus leads to the non-conservative nature of the
stress and a series of phenomena not found in classical elasticity, such as the
modification of defect strains, interactions and
motility~\cite{Scheibner2020,Braverman2021,Bililign2022}, and the emergence of
non-Hermitian skin effect~\cite{Chen2021} and chiral edge
waves~\cite{fossati2022}.

In the continuum description of odd elasticity, the odd-elastic modulus can be
obtained by the coarse-graining procedure of the non-conservative forces between
constituents in the elastic
body~\cite{Scheibner2020,Braverman2021,Poncet2022}.  Experimental
realizations of non-conservative interparticle forces include fluid-mediated
spinning particle~\cite{happel1983low,Sebastian2011}, gyroscopic
lattices~\cite{Brun2012,Lisa2015}, vortices in
superfluids~\cite{Tkachenko1969,Dung2020} and
skyrmions~\cite{Ochoa2017,Benzoni2021}. In previous studies, the
phenomenon of odd elasticity is attributed the active components of the system.
Theoretically, an antisymmetric modulus is introduced to successfully describe the
odd-elasticity behaviors at the continuum level. Specifically, a common approach to investigating odd elasticity begins with
the dynamic equation and the constitutive
relation~\cite{Salbreux2017,Scheibner2020,fruchart2022}, where an antisymmetric
odd-elastic modulus is introduced as a parameter. 
However, the origin of the
antisymmetry of the odd-elastic modulus in general has not yet been fully
understood. Exploring this fundamental scientific question yields deeper insights into
the nature of odd elasticity, and it also has strong connection to the
design of odd elastic systems. One challenge is that due to the
nonzero curl, the non-conservative stress involved in odd elasticity is not
derivable from a scalar potential like in classical elasticity.

In this work, a field theory in Hamiltonian formalism is constructed to produce
the antisymmetric elastic modulus tensor that is essential for a host of odd
elastic behaviors. Specifically, a $d$-dimensional continuum elastic body is
modeled as a Riemannian manifold embedded in the $(d+1)$-dimensional Euclidean
space, and the Hamiltonian for the elastic body of finite strain is constructed.
The key is introducing an anisotropic tensorial effective mass in the kinetic
energy term, as inspired by the work on Hamiltonian curl
forces~\cite{Berry2012,Berry2013,Berry2015}. The resulting antisymmetric elastic
modulus tensor also simultaneously inherits the anisotropic nature of the mass
tensor. Note that the inherent anisotropy is 
distinct from the odd elastic modulus generated through other mechanisms ~\cite{Scheibner2020,fruchart2022}.
The constitutive relation associated with the non-conservative stress is
nonlinear in general, and the nonlinearity originates from the intrinsic
geometry of the deformed elastic body. This work provides insights into the
origin of the antisymmetry of the elastic modulus tensor in odd elasticity,
and opens the promising possibility of exploring the dynamics of odd elasticity
in the Hamiltonian framework.

\section{Model and Method}

\subsection{Geometric viewpoint of elastic deformation}
An elastic body in continuum limit is modeled
as a $d$-dimensional Riemannian manifold $(\mathcal{B},g_{ab})$ embedded in
Euclidean space $\mathcal{E}$ in the study of its interior elastic deformation. 
The strain state is characterized by the metric tensor $g_{ab}$ defined on
$\mathcal{B}$.  For example, the strain-free elastic
body prior to any deformation is described by a Riemannian manifold
isometrically embedded in Euclidean space; the value of $g_{ab}$ is specified by
the pullback of the Euclidean metric~\cite{Efrati2009,Kupferman2017}. Note that
in this work we employ the abstract index notation in Latin letters to represent
a tensor; the tensor components are labeled by Greek letters.

\begin{figure}[!ht]
    \centering
    \includegraphics[width=0.5\linewidth]{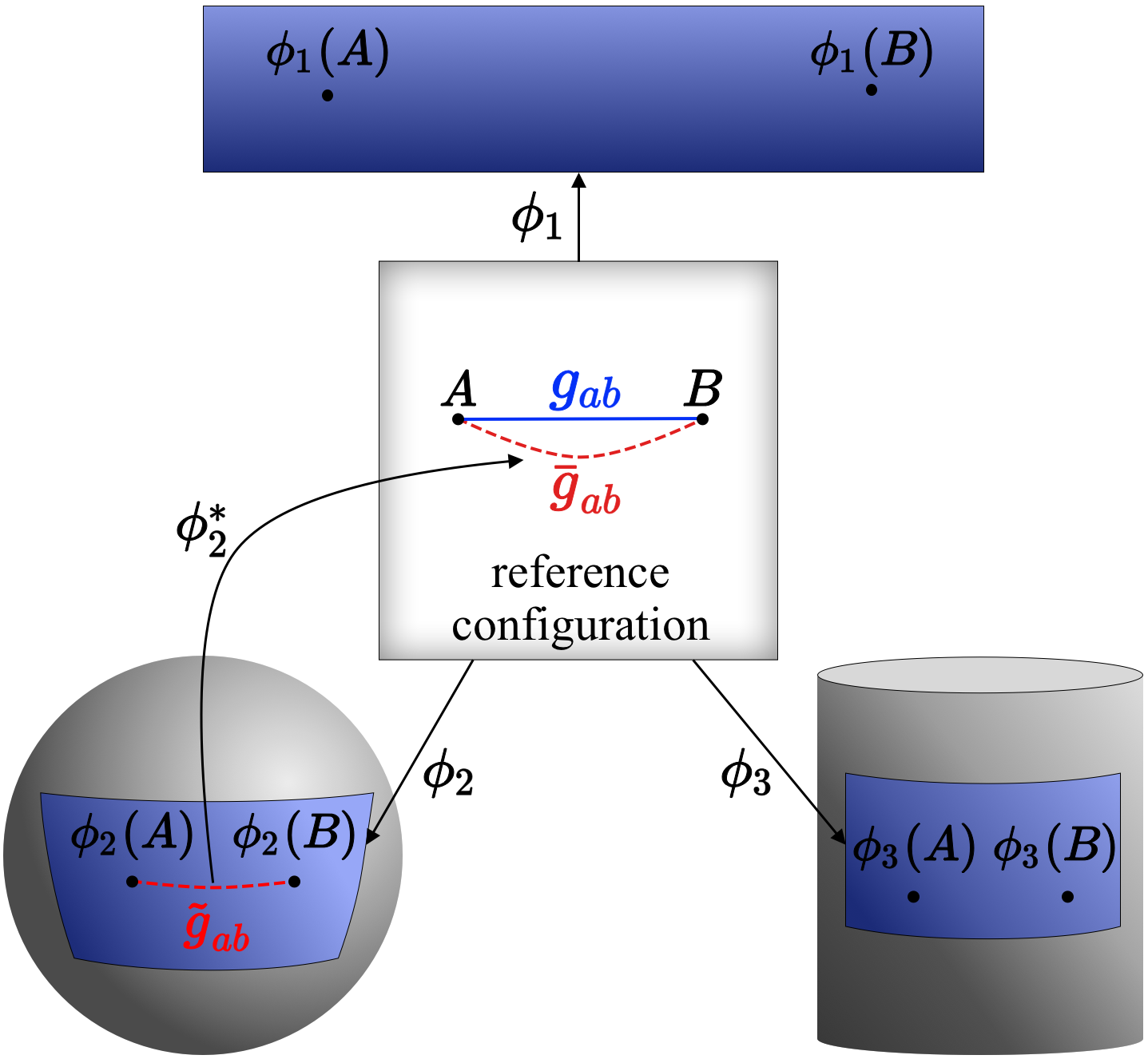}
    \caption{Illustration of the $\phi_i$
    mappings that connect the undeformed elastic body (represented by the
    central square as the reference configuration) to the various deformed
    configurations (represented by the
    surrounding figures). On the undeformed elastic body as the reference
    configuration, the length of the line element 
    connecting points $A$ and $B$ is determined by the metric $g_{ab}$. Taking the case of
    $\phi_2$ mapping (the left lower figure) for example, the reference line in
    solid blue is deformed to the dashed red line on the sphere, whose length is
    determined by the new metric $\tilde{g}_{ab}$. The corresponding metric
    $\bar{g}_{ab}$ on the reference configuration is connected to the deformed
    length via the pullback of $\tilde{g}_{ab}$: $\bar{g}_{ab} \equiv
    \phi^*\tilde{g}_{ab}$.    
  } \label{fig1}
\end{figure}

In general, the deformation of the body leads to the variation of the element of
length, and thus the metric of the manifold~\cite{Noll1978,Rougee1992}.
Therefore, the elastic deformation of the body could be characterized by a
diffeomorphism $\phi$ from some reference configuration to the deformed one:
$(\mathcal{B},g_{ab})\rightarrow(\mathcal{B},\tilde{g}_{ab})$.
Note that subscripts `$ab$' are to indicate that $g_{ab}$ is a
covariant tensor field of order 2.
The topology of
$\mathcal{B}$ is preserved in elastic deformation. To illustrate the $\phi$
mapping, we present some examples in figure~\ref{fig1}. Under the deformations
as described by the mappings $\phi_i$ ($i=1,2,3$), any given point in the
reference configuration, say $A$, is mapped to $\phi_i(A)$. The original 2D
elastic body of square shape is deformed to a rectangle (by $\phi_1$ mapping),
to the patches on the sphere (by $\phi_2$ mapping) and on the cylinder (by
$\phi_3$ mapping), respectively.  In these three kinds of deformations,
$\tilde{g}_{ab}$ refers to the Euclidean metric, the spherical metric and the
cylindrical metric, respectively.

From the geometric perspective, the deformation is characterized by the
variation of the metric from $g_{ab}$ to $\phi^*\tilde{g}_{ab}$, both of which
are defined on the same manifold $\mathcal{B}$. Specifically, the information
of the strain in the deformation is encoded in the difference between
$\phi^*\tilde{g}_{ab}$ and $g_{ab}$. To illustrate this point, let us consider
a curve $\gamma(t)\in(\mathcal{B},g_{ab})$, where $t\in[0,1]$. The length of
the line element on $\gamma$ is measured by $g_{ab}$. After the deformation, the
curve is mapped to $\phi(\gamma(t))\in(\mathcal{B},\tilde{g}_{ab})$. The length
of a line element on $\phi(\gamma)$ measured by $\tilde{g}_{ab}$ is equal to
the length of the corresponding line element on $\gamma$
measured by $\phi^*\tilde{g}_{ab}$. Therefore, the change of the curve length
could be measured by $\phi^*\tilde{g}_{ab}-g_{ab}$. As such, it is natural to
define the strain tensor on $\mathcal{B}$ as~\cite{Bilby1955,Landau1960,Kochetov1999}
\begin{eqnarray}
    u_{ab}=\frac{1}{2} \left(\bar{g}_{ab}-g_{ab}\right), \label{def_strain}
\end{eqnarray}
where $\bar{g}_{ab} \equiv \phi^*\tilde{g}_{ab}$. The definition of strain in
(\ref{def_strain}) is also applicable to large deformation~\cite{Landau1960}.

In the following discussion, the Riemannian manifold
$(\mathcal{B},\bar{g}_{ab})$ is denoted as $\Sigma$, representing the deformed
configuration of the elastic body. The elastic body is initially free of
stress, and is thus isometrically embedded in the $(d+1)$-dimensional
Euclidean space $\mathcal{E}$. The metric $\bar{g}_{ab}$ on the manifold
$\Sigma$ is induced from the metric $\delta_{ab}$ on
$\mathcal{E}$~\cite{Kupferman2017,Segev2020,Kolev2021}, i.e.,
$\bar{g}_{ab}=\delta_{ab}-N_aN_b$, where $N_a$ is the unit normal covector on
$\Sigma$ in $\mathcal{E}$.  For example, consider a 2D surface isometrically
embedded in $\mathcal{E}$. One may construct the Cartesian coordinates $\{X^1,
X^2, X^3\}$ centered at any point $P$ on the surface; $X^3$-axis is
perpendicular to the tangent plane at point $P$.  The first fundamental form
(or the line element) is given explicitly by the metric: $\bar{g}_{\alpha\beta}
(\mathrm{d} X^{\alpha})_a (\mathrm{d} X^{\beta})_b = \bar{g}_{\mu\nu} (\mathrm{d} s^{\mu})_a (\mathrm{d}
s^{\nu})_b + (\mathrm{d} X^3)_a(\mathrm{d} X^3)_b - (\mathrm{d} X^3)_a(\mathrm{d} X^3)_b = \bar{g}_{\mu\nu}
(\mathrm{d} s^{\mu})_a (\mathrm{d} s^{\nu})_b $, where $\{s^1 ,s^2 \}$ are local coordinates
near the point $P$.
Note that the Greek letters in the indices represent the
components of the tensor, and they take the values from $1$ to $d+1$. Here, the
convention of Einstein summation is applied.
For spherical surface, $s^{1}=\theta$ and $s^{2}=\phi$,
where the polar angle $\theta\in [0,\pi]$ and the azimuthal angle $\phi\in [0,
2\pi)$. Note that due to the isometric embedding of $\Sigma$ in $\mathcal{E}$,
all of the tensors in this work are regarded as being defined on the tangent
space of $\mathcal{E}$, and the lowering (or raising) of the indices of a tensor is
uniformly implemented by $\delta_{ab}$ (or $\delta^{ab}$).

Now, we characterize the displacement field associated with the deformation in terms
of the geometric language introduced in the preceding paragraph. We
first establish the Cartesian coordinates system of $\mathcal{E}$ by
$\{X^1, \dots , X^{d+1}\}$. The manifolds $(\mathcal{B},g_{ab})$
and $\Sigma$ are thus represented by the $\mathbb{R}^{d+1}$
-valued functions $X_0^{\mu}(p)$ and $X^{\mu}(\phi(p))$ ($\forall p \in
\mathcal{B}$), respectively. $\bar{X}^{\mu}(p) \equiv X^{\mu}(\phi(p))$. 
The displacement field on $\mathcal{B}$ is
\begin{eqnarray}
    U^a = U^{\mu} \left(\frac{\partial}{\partial X^{\mu}}\right)^a, \label{displacement_vector}
\end{eqnarray}
where the $\mu$-component of the displacement field $U^{\mu}(p)=
\bar{X}^{\mu}(p)-X_0^{\mu}(p), \forall p \in \mathcal{B}.$\\

\subsection{Construct Hamiltonian of the elastic body} The displacement fields $U^a$ constitute an infinite dimensional manifold called the
configuration space. The Hamiltonian of the elastic body is a scalar on the
cotangent bundle of the configuration space, and it can be written formally
as~\cite{Kolev2021}:
\begin{eqnarray}
    H = T + W = \int_{\Sigma}\rho\left(\mathscr{T}+\mathscr{W}\right)\bm{\varepsilon}, \label{H}
\end{eqnarray}
where $\mathscr{T}$ and $\mathscr{W}$ are the kinetic energy and elastic
potential per unit mass, respectively. $\bm{\varepsilon}$ is the Riemannian volume
form compatible with $\bar{g}_{ab}$. $\rho$ is the mass density of the deformed
configuration. In this work, $\mathscr{W}$ is a quadratic
local function of $u_{ab}$. The gradient of $\mathscr{W}$ is recognized as the
Kirchhoff stress tensor:
\begin{eqnarray}
    \tau^{ab} = \frac{\partial\mathscr{W}}{\partial u_{ab}}.\label{Kirchhoff_stress}
\end{eqnarray}
$\tau^{ab}$ is a symmetric tensor field on $\Sigma$. It
measures the stress experienced by the unit mass of the deformed
configuration. Note that the Cauchy stress tensor $\sigma^{ab}$ is
related to the Kirchhoff stress tensor by $\sigma^{ab} = \rho\tau^{ab}$, and it
measures the stress on the unit volume of the deformed
configuration~\cite{Kolev2021}. In
(\ref{H}), $\mathscr{T}$ could be written as a quadratic function of the
generalized momentum density field $P_a$ that is conjugated to $U^a$:
\begin{eqnarray}
    \mathscr{T} = \frac{1}{2}\mathscr{M}^{ab} P_a P_b. \label{kinetic_energy_density}
\end{eqnarray}

It is important to point out that here we introduce the effective mass tensor
$\mathscr{M}^{ab}$. $\mathscr{M}^{ab}$ is anisotropic and symmetric, and it is
invariant in the deformation of the elastic body. In connection
to possible experimental realizations, the anisotropy of
$\mathscr{M}^{ab}$ may be introduced by designing a 
local resonance cavity structure composed of an internal mass that is connected
to the wall by two perpendicular Hookean springs; such anisotropic mechanical
microstructures have been used to regulate the propagation of acoustic waves in
metamaterials~\cite{Liu2000,Huang2011}. An explanatory example based on the
modified mass-in-mass model is presented in section \ref{sec3.4} to show the realization
of anisotropic effective mass units. The idea of introducing the anisotropy in
the construction of the kinetic energy is inspired by the work on Hamiltonian
curl forces~\cite{Berry2012,Berry2013,Berry2015}. The concept of inducing
Hamiltonian curl forces through anisotropic mass was initially introduced in
systems with finite degrees of freedom by Berry and
Shukla~\cite{Berry2012,Berry2013,Berry2015}.  Specifically, it has been proved
that a class of non-conservative (i.e., whose curl is not zero) position
depending forces can be generated by the element of anisotropy in the kinetic
energy term in Hamiltonian. Here, we extend this idea to the continuum elastic
system by incorporating the non-conservative nature into $\mathscr{M}^{ab}$.
Consequently, (\ref{H}) admits a non-conservative force density, which is
crucial for understanding odd elasticity.

\section{RESULTS AND DISCUSSION}

\subsection{Hamilton's Equations}
The Hamilton's equations based on (\ref{H}) are 
\numparts
\begin{eqnarray}
   \frac{\partial U^a}{\partial t} = \frac{\delta H}{\delta P_a} = \rho\sqrt{\bar{g}}\mathscr{M}^{ab} P_b, \label{HE_u}\\
   \frac{\partial P_a}{\partial t} = -\frac{\delta H}{\delta U^a} = \sqrt{\bar{g}} f_a, \label{HE_p}
\end{eqnarray}
\endnumparts
where $f_a = \bar{g}_b{}^c\partial_c(\bar{g}_{ad}\sigma^{db})$.
Note that the repeated indices in pairs represents tensor
contraction.
$\bar{g}$ is the determinant of $\bar{g}_{ab}$;
$\partial_c$ is the Cartesian derivative operator on $\mathcal{E}$ acting on the
embedding coordinates $\{\bar{X}^\mu\}$ of $\Sigma$. The detailed
information is presented in \ref{AA} and \ref{AB}. Note that
the generalized coordinates $U^a$ in the Hamiltonian formalism is closely
related to the strain of an elastic body.

By combining (\ref{HE_u}) and (\ref{HE_p}), we obtain the dynamic
equation of the manifold $\Sigma$ as
\begin{eqnarray}
    \frac{\partial V^a}{\partial t} = M^{ab}f_b \equiv F^a,\label{dynamics_equation}
\end{eqnarray}
where the displacement velocity field $V^a \equiv \partial{U^a}/\partial{t}$ and
the symmetric tensor field $M^{ab} \equiv \rho\bar{g}\mathscr{M}^{ab}$.
According to (\ref{dynamics_equation}), $F^{a}$ regulates
the acceleration of unit mass in the elastic body, and it is recognized as the
Newtonian force per unit mass. (\ref{dynamics_equation}) suggests
that the dynamics of a continuum elastic body with anisotropic mass $\mathscr{M}^{ab}$
under $f^a$ is equivalent to the dynamics of a unit scalar mass
under $F^a$; the velocity $V^a$ as associated to the displacement field is the
common dynamics variable.

\subsection{Hamiltonian curl forces and odd elasticity}
Without any loss of generality, $M^{ab}$ is decomposed as
\begin{eqnarray}
  M^{ab} = M\delta^{ab} + \Omega^{ab},\label{Mab}
\end{eqnarray}
where $M = \delta_{ab}M^{ab}/(d+1)$, and $\Omega^{ab}$ is a traceless
symmetric tensor field. Here, we shall emphasize that the anisotropic effect
in the kinetic energy is captured by $\Omega^{ab}$. By (\ref{Mab}),
the kinetic energy density $\mathscr{T}$ is divided into the
isotropic and the anisotropic parts:
\begin{eqnarray}
  \mathscr{T} = \frac{1}{2\rho^2\bar{g}}P_aP^a +
  \frac{1}{2\rho\bar{g}}\Omega^{ab} P_aP_b. \label{separated_T}
\end{eqnarray}
We also cast $F^a$ in the following form
\begin{eqnarray}
    F^{a} = M f^a + \Omega^{ab}f_b. \label{anisotropic_force}
\end{eqnarray}

Now, let us focus on (\ref{anisotropic_force}).
The first term in (\ref{anisotropic_force}) is recognized as the force per
unit mass as a derivative of the elastic potential for
$M=1/\rho$~\cite{Kochetov1999,Guven2002,Guven2004a,Guven_2004}. 
According
to (\ref{HE_u}) and (\ref{HE_p}), the power of $M f^a$
is expressed as:
\begin{eqnarray}
  M f_aV^a=\mathscr{M}^{ab}\frac{\partial P_a}{\partial t}P_b=\frac{\mathrm{d}\mathscr{T}}{\mathrm{d}t}=-\frac{\mathrm{d}\mathscr{W}}{\mathrm{d}t}.\label{power_fa}
\end{eqnarray}
In the last equality, the conservation of the total energy is applied, i.e.,
$\mathrm{d}{H}/\mathrm{d}{t}=0$. According to (\ref{power_fa}), the power of $M
f^a$ is equal to the reduction rate of $\mathscr{W}$. Due to the conservation of
the total energy, it indicates that the work done by $M f^a$ is equal to the
energy transfer from the potential energy to the kinetic energy per unit mass in
the deformation of the elastic body. Note that for a single-valued $\mathscr{W}$
function, the integration of (\ref{power_fa}) over any closed orbits in
$\mathcal{E}$ is zero, implying that $M f^a$ is a conservative force. As such,
the first term of $F^{a}$ in (\ref{anisotropic_force}) does not alter the
value of either $\mathscr{W}$ or $\mathscr{T}$ in a cyclic deformation of
the elastic body.

We proceed to examine the second term $\Omega^{ab} f_b$ in (\ref{anisotropic_force}).
The power of $\Omega^{ab} f_b$ is:
\begin{eqnarray}
  \Omega^{ab} f_b V_a &= \rho\Omega^{ab}\frac{\partial P_b}{\partial t}\mathscr{M}_a{}^cP_c \nonumber\\
  &= \frac{1}{\rho
  \bar{g}}\Omega^{ab}\left(\delta_a{}^c+\rho\Omega_a{}^c\right)P_c\frac{\partial P_b}{\partial t}.\label{Oab}
\end{eqnarray}
Under the assumption that the anisotropic tensor $\rho \Omega^{ab}$ is small compared
with $\delta^{ab}$, (\ref{Oab}) becomes
\begin{eqnarray}
\Omega^{ab} f_b V_a = \frac{\mathrm{d}}{\mathrm{d}t}\left(\frac{1}{2\rho\bar{g}}\Omega^{ab} P_aP_b\right).\label{power_ofa}
\end{eqnarray}
Here, it is important to point out that the R.H.S. of (\ref{power_ofa}) is
identified as the rate of the increase of the anisotropic part
of $\mathscr{T}$. In other words, the work done by $\Omega^{ab} f_b$ causes 
the increase in the anisotropic part of the kinetic energy per unit mass.

In the following, we will show that the second term $\Omega^{ab} f_b$ in
(\ref{anisotropic_force}) is a non-conservative force. To this end, we first 
substitute the expression of $f_a$ in
(\ref{anisotropic_force}), and obtain the expression for the Newtonian force per unit volume as
\begin{eqnarray}
  \rho F^{a} &= \bar{g}_b{}^c\partial_c\sigma^{ab} +
  \rho\Omega^{ad}\bar{g}_b{}^c\partial_c\left(\bar{g}_{de}\sigma^{eb}\right)\nonumber\\
  &= \bar{g}_b{}^c\partial_c\left(\sigma^{ab}+\hat{\sigma}^{ab}\right)
    - \bar{g}_{bd}\sigma^{dc}\partial_c\left(\rho\Omega^{ab}\right), \label{intrinsic_omega}
\end{eqnarray}
where $\hat{\sigma}^{ab} \equiv \rho\Omega^{ac}\bar{g}_{cd}\sigma^{db}$. Note
that $\hat{\sigma}^{ab}$ is not symmetric. Furthermore, $\hat{\sigma}^{ab}$ does
not live in the tangent space of $\Sigma$ because of the acting of
$\Omega^{ac}\bar{g}_{cd}$. For spatially-slowly-varying $\rho\Omega^{ab}$, its
gradient could be regarded as a small quantity.  Therefore, the last term in
(\ref{intrinsic_omega}) is much smaller than the second term; note that
$\bar{g}_b{}^c\partial_c \hat{\sigma}^{ab} = \bar{g}_b{}^c\partial_c
(\rho\Omega^{ad}\bar{g}_{de}\sigma^{eb})$.  As such, the last term could be
neglected.  The anisotropic effect boils down to the modification of
$\hat{\sigma}^{ab}$ in $\sigma^{ab}$ as shown in the first bracket in
(\ref{intrinsic_omega}).

Now, to reveal the non-conservative nature of the second term in
(\ref{anisotropic_force}), we analyze the work done by the total stress per
unit mass in a cyclic deformation. In the
deformation process, the instantaneous strain state of the elastic body is
denoted by $u_{ab}(t)$, which is represented by a point in the space of the
strain field $u_{ab}$. The cycle of deformation is thus described by a loop
$\eta(t)$ in the strain space. $t\in[0, T]$. $\eta(0)=\eta(T)$. The work done per
unit mass in a cycle of deformation is~\cite{Scheibner2020,fossati2022}:
\begin{eqnarray}
    \Delta \mathscr{A} &= \oint_{\eta(t)}\left(\tau^{ab}+\hat{\tau}^{ab}\right)\mathrm{d} u_{ab}\nonumber\\
    &= \int_S\left(\frac{\partial \tau^{ab}}{\partial u_{cd}} +
    \frac{\partial \hat{\tau}^{ab}}{\partial u_{cd}}\right)\mathrm{d} u_{cd}\wedge\mathrm{d} u_{ab}, \nonumber \\
    &\equiv \int_S T^{abcd} \mathrm{d} u_{cd}\wedge\mathrm{d} u_{ab}, \label{dW}
\end{eqnarray}
where $\hat{\tau}^{ab} = \hat{\sigma}^{ab}/\rho$, and $S$ is the surface
enclosed by $\eta(t)$. Stokes's theorem is used in the derivation for
(\ref{dW}). Since the exterior product is antisymmetric, the stress is
conservative (such that $\Delta \mathscr{A}=0$ in the cyclic deformation) if
and only if $T^{abcd}$ possesses the major index symmetry, i.e., $T^{abcd} =
T^{cdab}$.  Especially, quasi-static cyclic deformation can be realized by
applying suitable external force on $\Sigma$. $\Delta \mathscr{A}$ is then
equal to the work done by the external force.

We analyze the first term of $T^{abcd}$ in (\ref{dW}).
From the expression for $\mathscr{W}$
\begin{eqnarray}
    \mathscr{W} = \frac12 C^{abcd} u_{ab}u_{cd},\label{quadratic_w}
\end{eqnarray}
where $C^{abcd}$ is the elastic modulus tensor with major index symmetry
(i.e, $C^{abcd}=C^{cdab}$), we obtain the linear constitutive relation
\begin{eqnarray}
    \tau^{ab} = C^{abcd} u_{cd}.\label{hookes_law}
\end{eqnarray}
Due to the major index symmetry of $C^{abcd}$, the first term of $T^{abcd}$
satisfies $\partial{\tau^{ab}}/\partial{u_{cd}}=\partial{\tau^{cd}}/\partial{u_{ab}}$, which indicates
the conservative nature of $\tau^{ab}$. In other words, the work done by
$\tau^{ab}$ is zero in a cyclic deformation.

For the second term in $T^{abcd}$, from the definition of $\hat{\sigma}^{ab}$
and (\ref{hookes_law}), we have 
\begin{eqnarray}
   \hat{\tau}^{ab} &=& \rho\Omega^{ac}\bar{g}_{cd}\tau^{db} \nonumber \\ 
   &=& \rho\Omega^{ae} \left(g_{ef} + 2u_{ef}\right) C^{fbcd} u_{cd}  \nonumber \\
  &=& \hat{C}^{abcd}u_{cd} + \hat{D}^{abcdef} u_{cd}u_{ef},\label{odd_consre}
\end{eqnarray}
where
\begin{eqnarray}
  \hat{C}^{abcd} = \rho\Omega^{ae} g_{ef} C^{fbcd},\quad 
  \hat{D}^{abcdef}=2\rho\Omega^{ac} C^{dbef}. \label{Chat}
\end{eqnarray}
In the nonlinear constitutive relation in (\ref{odd_consre}), the quadratic
term naturally arises in the expansion of $\bar{g}_{cd}$ as the sum of the
metric $g_{cd}$ of the reference configuration and the strain field
$2u_{cd}$.  Furthermore, the anisotropic tensor $\Omega^{ab}$
leads to the anisotropic nature of the elastic modulus $\hat{C}^{abcd}$ and
$\hat{D}^{abcdef}$.  Here, we shall point out that in the Hamiltonian formalism
based on anisotropic mass tensor, the resulting odd elastic modulus are
naturally anisotropic.  The anisotropy of $\hat{C}^{abcd}$ and
$\hat{D}^{abcdef}$ is inherent in our formalism. A detailed discussion about the
symmetry of $\hat{C}^{abcd}$ and $\hat{D}^{abcdef}$ is in \ref{AD}.
In general, the odd elastic modulus are not necessarily anisotropic in
2D~\cite{Scheibner2020}.  By (\ref{odd_consre}), the second term of
$T^{abcd}$ is obtained
\begin{eqnarray}
  \frac{\partial \hat{\tau}^{ab}}{\partial u_{cd}} = \hat{C}^{abcd} + \left(\hat{D}^{abcdef} + \hat{D}^{abef(cd)}\right) u_{ef}. \label{T_2}
\end{eqnarray}
The bracket in the superscript of $\hat{D}$ indicates the symmetric part of the
tensor.
Here, it is important to point out that in (\ref{Chat}), the
involvement of the $\Omega^{ab}$-tensor breaks the major index symmetry of
$\hat{C}^{abcd}$, as well as that of the second and the third terms in the right
hand side of (\ref{T_2}). Consequently, $\hat{\tau}^{ab}$ is
non-conservative according to (\ref{T_2}).

Note that both $\hat{C}^{abcd}$ and $\hat{D}^{abcdef}$ in (\ref{Chat}) are
invariant in the deformation of $\Sigma$. The reason is as follows. In the
expressions for $\hat{C}^{abcd}$ and $\hat{D}^{abcdef}$, both the elastic
modulus tensor $C^{abcd}$ and the metric $g_{ab}$ are independent of the
deformation of $\Sigma$. Regarding the factor $\rho\Omega^{ab}$,  according to
(\ref{Mab}) and the definition for $M^{ab}$,
$\rho\Omega^{ab}=\rho^2\bar{g}\mathscr{M}^{ab}-\delta^{ab} = \rho_0^2
g\mathscr{M}^{ab}-\delta^{ab}$, where $\rho_0$ is the mass density of the
reference configuration. $\rho\Omega^{ab}$ is therefore invariant in the
deformation of $\Sigma$.

In the regime of small deformation, where the quadratic term in
(\ref{odd_consre}) can be neglected, the constitutive relation of
$\hat{\tau}^{ab}$ becomes linear:
\begin{eqnarray}
    \hat{\tau}^{ab} = \hat{C}^{abcd}u_{cd}.\label{linear_odd}
\end{eqnarray}
The broken major index symmetry of $\hat{C}^{abcd}$ indicates the presence of the
antisymmetric (odd) part in the elastic modulus tensor, which is responsible for
the non-conservativity of the stress and the extra work occurring in cyclic
deformations. As such, $\hat{C}^{abcd}$ is called the odd elastic modulus in
literature~\cite{Scheibner2020}. Microscopic mechanism for odd elasticity has
been attributed to various non-conservative interparticle forces.  Here, the odd
elasticity as characterized by the $\hat{C}^{abcd}$-tensor in
(\ref{linear_odd}) is derived from the anisotropic effective mass in the
Hamiltonian formalism in the continuum level. \\

We shall point out that in the model of odd elasticity based on the
Hamiltonian formalism, the total energy is conserved. The work extracted by
$\hat{\tau}^{ab}$ during cyclic deformations is from the conversion of the
kinetic energy pre-stored in the internal degree of freedom of the system,
rather than from the external energy inputs (see section \ref{sec3.4} for detailed
discussions).  The anisotropy of the effective mass
$\mathscr{M}^{ab}$ arises from the coarse-grained internal degrees of freedom.
In a cyclic process, according to (\ref{power_ofa}), the non-conservative
curl force transfers the kinetic energy between the (coarse-grained) anisotropic
parts of $\mathscr{T}$ and the isotropic parts of $\mathscr{T}$. As such, for
a cyclic deformation starting and ending at zero velocity, no work can be
extracted.

\subsection{Example of 2D planar deformation}
In this section, we illustrate the emergence of odd elasticity for the simple
case of planar deformation of a 2D elastic body under small deformation approximation.

The strain tensor $u_{ab}$ in the deformation can be expanded as 
$u_{ab}=u_\mu\mathfrak{e}^\mu{}_{ab}$, where $\mathfrak{e}^\mu{}_{ab}$ is a set
of orthogonal tensor bases on the reference configuration equipped with a local
orthonormal frame fields $\{e_1{}^a,e_2{}^a\}$~\cite{Scheibner2020}:
\numparts
\begin{eqnarray}
    \mathfrak{e}^1{}_{ab}=e^1{}_a e^1{}_b + e^2{}_a e^2{}_b,\\
    \mathfrak{e}^2{}_{ab}=e^2{}_a e^1{}_b - e^1{}_a e^2{}_b, \\
    \mathfrak{e}^3{}_{ab}=e^1{}_a e^1{}_b - e^2{}_a e^2{}_b, \\
    \mathfrak{e}^4{}_{ab}=e^1{}_a e^2{}_b + e^2{}_a e^1{}_b.\label{tensor_frame}
\end{eqnarray}
\endnumparts
$u_\mu$ characterizes the following local modes of deformation: dilation,
rotation, and shearing.  The stress tensor can also be expanded as $\tau^{ab} =
\tau^\mu\mathfrak{e}_\mu{}^{ab}$, where $\tau^\mu$ are associated with pressure,
torque density and shear stress.

The rotationally symmetric elastic modulus tensor field $C^{abcd}$ could be
written as (see \ref{AC} for details):
\begin{eqnarray}
    C^{abcd}=\lambda g^{ab}g^{cd}+2\mu g^{a(c}g^{d)b},\label{C}
\end{eqnarray}
where $\lambda$ and $\mu$ are Lam\'e coefficients. In the tensor bases
of $\{ \mathfrak{e}^\mu{}_{ab} \}$, $g_{ab}=\mathfrak{e}^1{}_{ab}$, and
$C^{abcd} = C^{\mu\nu}\mathfrak{e}_\mu{}^{ab}\mathfrak{e}_\nu{}^{cd}$. From
(\ref{C}), we obtain
\begin{eqnarray}
    C^{\mu\nu} =
    \begin{bmatrix}
        \lambda+\mu & 0 & 0 & 0\\
        0 & 0 & 0 & 0\\
        0 & 0 & \mu & 0\\
        0 & 0 & 0 & \mu\\
    \end{bmatrix}. \label{C_munu}
\end{eqnarray}
Correspondingly, (\ref{hookes_law}) is expressed as $\tau^\mu=C^{\mu\nu}u_\nu$.
$\lambda+\mu$ and $\mu$ characterize the isotropic stretching rigidity
and the isotropic shear rigidity, respectively. Note that the matrix in \ref{C_munu}
is invariant under the rotation of $\{e_1{}^a,e_2{}^a\}$.

We proceed to discuss the components $\hat{C}^{\mu\nu}$. 
By expanding the traceless tensor $\rho\Omega^{ab}$ in the
tensor bases of $\{e_1{}^a,e_2{}^a\}$ as
\begin{eqnarray}
   \rho \Omega^{ab} = \kappa e_1{}^a e_1{}^b + \zeta e_1{}^a e_2{}^b + \zeta e_2{}^a e_1{}^b - \kappa e_2{}^a e_2{}^b,\label{anisotropic_mass_component}
\end{eqnarray}
where $\kappa$ and $\zeta$ are scalars, we finally have 
\begin{eqnarray}
    \hat{C}^{\mu\nu} =
    \begin{bmatrix}
        0 & 0 & \kappa\mu & \zeta\mu \\
        0 & 0 & \zeta\mu & -\kappa\mu \\
        \kappa(\lambda+\mu) & 0 & 0 & 0 \\
        \zeta(\lambda+\mu) & 0 & 0 & 0
    \end{bmatrix}.\label{coordinates_hC}
\end{eqnarray}
Since $\hat{C}^{abcd} = \hat{C}^{\mu\nu}\mathfrak{e}_\mu{}^{ab}\mathfrak{e}_\nu{}^{cd}$,
the abstract index exchange $ab \leftrightarrow cd$ in $\hat{C}^{abcd}$ is identical
to the label exchange $\mu \leftrightarrow \nu$ in $\hat{C}^{\mu\nu}$.
(\ref{coordinates_hC}) shows that $\hat{C}^{\mu\nu} \neq
\hat{C}^{\nu\mu}$, and therefore $\hat{C}^{abcd} \neq \hat{C}^{cdab}$. This key feature of the broken major symmetry
is exactly the mathematical structure underlying the phenomenon of odd
elasticity. The upper right $2\times 2$ submatrix in (\ref{coordinates_hC})
connects shear strain with pressure and torque, and the lower left $2\times 2$
submatrix connects dilation with shear stress. The existence of these two
non-zero submatrices is an indicator for the anisotropy of
$\hat{C}^{abcd}$~\cite{Scheibner2020}.

\subsection{Example of an anisotropic mass model}\label{sec3.4}

In the following, we employ the modified mass-in-mass model as an example
to show the emergence of the anisotropic inertial mass and the non-conservative
force.

\begin{figure}[!ht]
  \centering
  \includegraphics[width=0.25\linewidth]{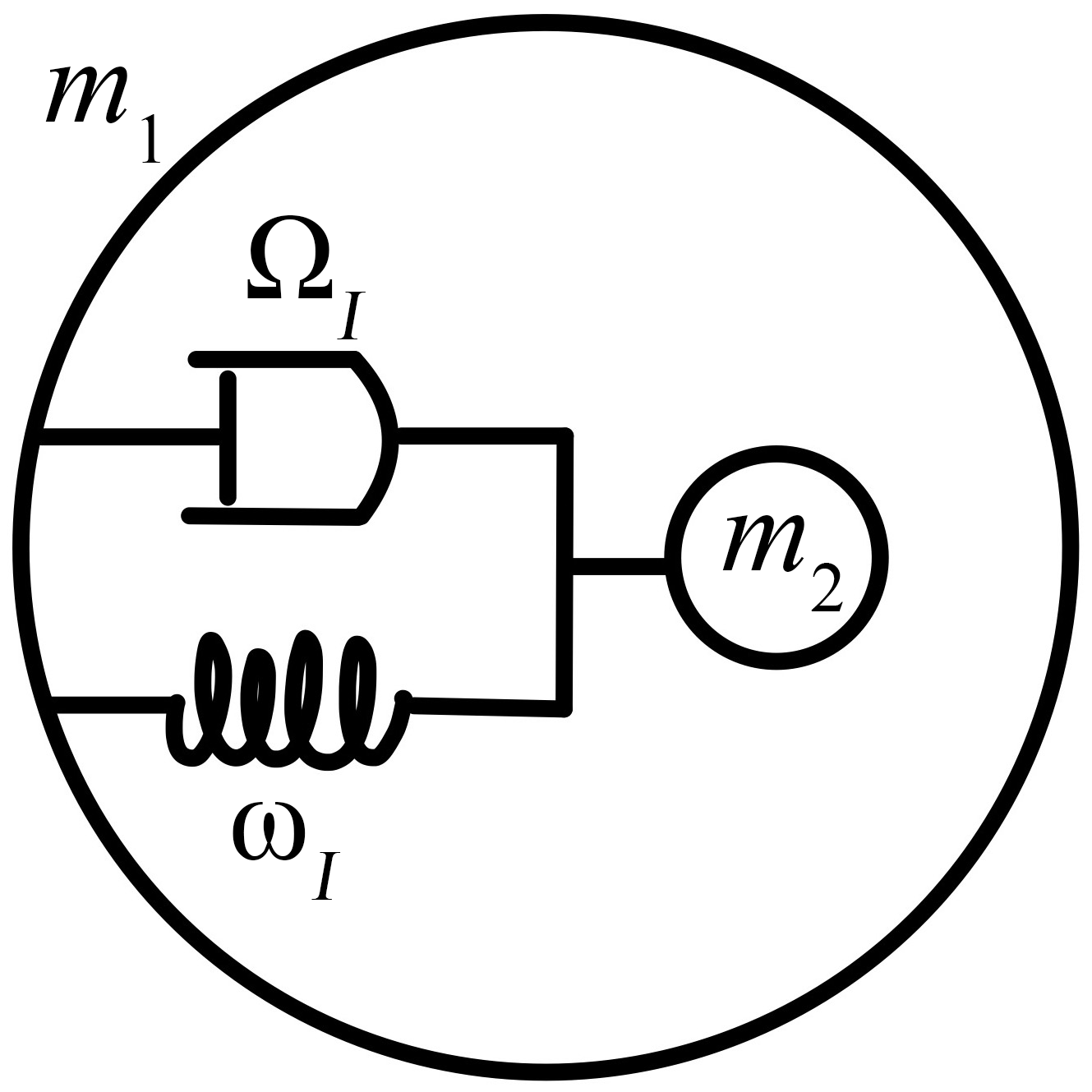}
  \caption{2D anisotropic mass model.} \label{fig3}
\end{figure}

Consider a 2D cavity of mass $m_1$ that contains an internal mass
$m_2$, as illustrated in figure~\ref{fig3}. The internal mass is connected to
the cavity via a spring of resonance frequency $\omega_\mathrm{I}$ and a damper
of damping constant $\Omega_\mathrm{I}$. The cavity is subject to an
external potential $W$. The dynamic equations for this mass-in-mass model are
\numparts
    \begin{eqnarray}
        m_1 \frac{\mathrm{d}^2 X_1}{\mathrm{d} t^2} = -\frac{\partial W}{\partial X_1} - m_2 \omega_\mathrm{I}^2 D - \Omega_\mathrm{I} \frac{\mathrm{d} D}{\mathrm{d} t},\label{eq1a}\\
        m_2 \frac{\mathrm{d}^2 X_2}{\mathrm{d} t^2} =  m_2 \omega_\mathrm{I}^2 D + \Omega_\mathrm{I} \frac{\mathrm{d} D}{\mathrm{d} t},\label{eq1b}\\
        m_1 \frac{\mathrm{d}^2 Y_1}{\mathrm{d} t^2} = -\frac{\partial W}{\partial Y_1},\label{eq1c}\\
        m_2 \frac{\mathrm{d}^2 Y_2}{\mathrm{d} t^2} = 0,\label{eq1d}
    \end{eqnarray}
\endnumparts
where $(X_1,\ Y_1)$ and $(X_2,\ Y_2)$ represent the Cartesian coordinates of
displacements of $m_1$ and $m_2$ respectively, and $D = X_1-X_2$.

We focus on the motion along the x-direction. Adding and subtracting
(\ref{eq1a}) and (\ref{eq1b}) lead to:
\numparts
    \begin{eqnarray}
       \left(m_1+m_2\right) \frac{\mathrm{d}^2 X_1}{\mathrm{d} \hat{t}^2} - m_2 \frac{\mathrm{d}^2 D}{\mathrm{d} \hat{t}^2}= f_x,\label{eq2a}\\
       m_1 \frac{\mathrm{d}^2 D}{\mathrm{d} \hat{t}^2} + \left(m_1+m_2\right) \left(\hat{\omega}^2 D + \hat{\Omega}\frac{\mathrm{d} D}{\mathrm{d} \hat{t}}\right) = f_x,\label{eq2b}
    \end{eqnarray}
\endnumparts
where the external force:
\begin{eqnarray}
    f_x=-\frac{1}{\omega_\mathrm{E}^2}\frac{\partial W}{\partial X_1},
\end{eqnarray}
and the dimensionless quantities are 
\begin{eqnarray}
    \hat{t}=t\omega_\mathrm{E},\quad \hat{\omega}=\frac{\omega_\mathrm{I}}{\omega_\mathrm{E}},\quad
    \hat{\Omega}=\frac{\Omega_\mathrm{I}}{m_2 \omega_\mathrm{E}}.
\end{eqnarray}
We focus on the regime of $\hat{\omega}\gg 1$ and $\hat{\Omega}\gg 1$, i.e., 
the dynamics of the internal degree of freedom is much
faster than the that of the cavity.

We first analyze (\ref{eq2b}) by introducing the following Green's function $G_\omega$:
\begin{eqnarray}
    D(\hat{t}) &= \frac{1}{2\pi}\int_{-\infty}^{+\infty}\mathrm{d} \omega \int_{-\infty}^{+\infty}\mathrm{d} \hat{t}' \exp[-\mathrm{i}\omega(\hat{t}-\hat{t}')]G_\omega f_x(\hat{t}')\nonumber\\
    &=  \frac{1}{2\pi}\int_{-\infty}^{+\infty}\mathrm{d} \omega \int_{-\infty}^{+\infty}\mathrm{d} \hat{t}' \exp(-\mathrm{i}\omega\hat{t}')G_\omega f_x(\hat{t}-\hat{t}'),\label{eq3}
\end{eqnarray}
where
\begin{eqnarray}
    G_\omega = \frac{1}{(m_1+m_2)\left(\hat{\omega}^2
    -\mathrm{i}\hat{\Omega}\omega\right) - m_1 \omega^2}.\label{eq4}
\end{eqnarray}
We thus have:
\begin{eqnarray}
    \frac{\mathrm{d}^2 D}{\mathrm{d} \hat{t}^2} = -\frac{1}{2\pi}\int_{-\infty}^{+\infty}\mathrm{d} \omega \int_{-\infty}^{+\infty}\mathrm{d} \hat{t}' \exp(-\mathrm{i}\omega\hat{t}') \omega^2 G_\omega f_x(\hat{t}-\hat{t}').\label{eq5}
\end{eqnarray}

The integration over $\omega$ in (\ref{eq5}) is implemented by the residue
theorem. Specifically, 
\begin{eqnarray}
    &\int_{-\infty}^{+\infty}\mathrm{d} \omega \exp(-\mathrm{i}\omega\hat{t}') \omega^2 G_\omega = \int_{-\infty}^{+\infty}\mathrm{d} \omega \frac{\exp(-\mathrm{i}\omega\hat{t}')\omega^2}{\left(m_1+m_2\right)\left(\hat{\omega}^2 -\mathrm{i}\hat{\Omega}\omega\right) - m_1 \omega^2}\nonumber\\
    &= \int_{-\infty}^{+\infty}\mathrm{d} \omega \frac{\exp(-\mathrm{i}\omega\hat{t}')\left(m_1+m_2\right)\left(\hat{\omega}^2 -\mathrm{i}\hat{\Omega}\omega\right)}{m_1\left(m_1+m_2\right)\left(\hat{\omega}^2 -\mathrm{i}\hat{\Omega}\omega\right) - m_1^2 \omega^2} - \frac{1}{m_1}\int_{-\infty}^{+\infty}\mathrm{d} \omega \exp(-\mathrm{i}\omega\hat{t}')\nonumber\\
    &= \int_{-\infty}^{+\infty}\mathrm{d} \omega \frac{\exp(-\mathrm{i}\omega\hat{t}')\left(m_1+m_2\right)\left(\mathrm{i}\hat{\Omega}\omega - \hat{\omega}^2\right)}{m_1^2\left(\omega-\omega_+\right)\left(\omega-\omega_-\right)} - \frac{2\pi}{m_1}\delta(\hat{t}')
    .\label{eq6}
\end{eqnarray}
The integrand has two poles in the lower half complex $\omega$-plane at:
\begin{eqnarray}
    \omega_{\pm} = \frac{-\mathrm{i}\left(m_1+m_2\right)\hat{\Omega}\pm \sqrt{\left(m_1+m_2\right) \left[\left(m_1+m_2\right)\hat{\Omega}^2-4m_1\hat{\omega}^2\right]}}{2m_1},\label{eq7}
\end{eqnarray}
if $\left(m_1+m_2\right)\hat{\Omega}^2>4m_1\hat{\omega}^2$.

By the standard method of constructing a closed contour via introducing an
infinitely large semicircle in the lower half-plane, and applying the residue
theorem, the first term in (\ref{eq6}) becomes
\begin{eqnarray}
    &-&\frac{1}{2\pi\mathrm{i}}\int_{-\infty}^{+\infty}\mathrm{d} \omega \frac{\exp(-\mathrm{i}\omega\hat{t}')\left(m_1+m_2\right)\left(\hat{\omega}^2 -\mathrm{i}\hat{\Omega}\omega\right)}{m_1\left(m_1+m_2\right)\left(\hat{\omega}^2 -\mathrm{i}\hat{\Omega}\omega\right) - m_1^2 \omega^2}\nonumber
    \\
    &=& \frac{\exp(-\mathrm{i}\omega_+\hat{t}')\left(m_1+m_2\right)\left(\mathrm{i}\hat{\Omega}\omega_+ - \hat{\omega}^2\right)}{m_1^2\left(\omega_+-\omega_-\right)}\nonumber\\
    &&+ \frac{\exp(-\mathrm{i}\omega_-\hat{t}')\left(m_1+m_2\right)\left(\mathrm{i}\hat{\Omega}\omega_- - \hat{\omega}^2\right)}{m_1^2\left(\omega_--\omega_+\right)}
    .\label{eq8}
\end{eqnarray}

By inserting (\ref{eq6}) and (\ref{eq8}) into (\ref{eq5}), we obtain:
\begin{eqnarray}
   \frac{\mathrm{d}^2 D}{\mathrm{d} \hat{t}^2} &=& \int_{0}^{+\infty}\mathrm{d}\hat{t}' \frac{\mathrm{i} \exp(-\mathrm{i}\omega_+\hat{t}')\left(m_1+m_2\right)\left(\mathrm{i}\hat{\Omega}\omega_+ - \hat{\omega}^2\right)}{m_1^2\left(\omega_+-\omega_-\right)} f_x(\hat{t}-\hat{t}')\nonumber\\
    &&+ \int_{0}^{+\infty}\mathrm{d}\hat{t}' \frac{\mathrm{i} \exp(-\mathrm{i}\omega_-\hat{t}')\left(m_1+m_2\right)\left(\mathrm{i}\hat{\Omega}\omega_- - \hat{\omega}^2\right)}{m_1^2\left(\omega_--\omega_+\right)} f_x(\hat{t}-\hat{t}') + \frac{f_x(\hat{t})}{m_1}\nonumber\\
    &=& -\frac{f_x(\hat{t})}{m_1} + \frac{f_x(\hat{t})}{m_1}\nonumber\\
    &=& 0.\label{eq9}
\end{eqnarray}
Note that the initial value theorem of Laplace transform is used in 
the integration over $\hat{t}'$. The theorem states that for large real $s$,
\begin{eqnarray}
    \int_{0}^{+\infty}\mathrm{d} \hat{t}' f(\hat{t}')\exp(-s \hat{t}') = \frac{f(0)}{s}.\label{eq10}
\end{eqnarray}

Consequently, (\ref{eq2a}) is simplified to:
\begin{eqnarray}
    \left(m_1+m_2\right) \frac{\mathrm{d}^2 X_1}{\mathrm{d} \hat{t}^2} = -\frac{1}{\omega_\mathrm{E}^2}\frac{\partial W}{\partial X_1}.\label{eq11}
\end{eqnarray}
Along with (\ref{eq1c}), the equations of motion for the displacement
$R^a=(X_1,\ Y_1)$ are finally reduced to:
\begin{eqnarray}
    m_{ab}\frac{\mathrm{d}^2 R^a}{\mathrm{d} \hat{t}^2} = -\nabla_a \hat{W},\label{eq12}
\end{eqnarray}
where $\hat{W}=W/\omega_\mathrm{E}^2$. Here, it is important to point out that the effective
anisotropic mass arises in the mass-in-mass model. In Cartesian coordinates, 
\begin{eqnarray}
    m_{\mu\nu} = \begin{bmatrix}
                    m_1+m_2 & 0\\
                    0 & m_1
                \end{bmatrix}.\label{eq13}
\end{eqnarray}
Therefore, elastic media fabricated by connecting such cavity units with Hookean
springs could manifest anisotropic mass density, and thus exhibit odd elasticity.

We proceed to discuss the Hamiltonian non-conservative force acting on the cavity.
First of all, we construct the Hamiltonian corresponding to
(\ref{eq12}) as
\begin{eqnarray}
    H = T + \hat{W}.\label{eq14}
\end{eqnarray}
The total kinetic energy $T$ includes the kinetic energies of $m_1$ and
$m_2$:
\begin{eqnarray}
    T = \frac12 M^{ab}P_aP_b = \frac12 m_1 \left(u^2 + v^2\right) + \frac12 m_2 u^2,\label{eq15}
\end{eqnarray}
where $M^{ab}=(m^{-1})^{ab}$, $u = \mathrm{d}{X_1}/\mathrm{d}{\hat{t}}$, $v=\mathrm{d}{Y_1}/\mathrm{d}{\hat{t}}$.
The Hamilton's equations are:
\numparts
    \begin{eqnarray}
        \frac{\mathrm{d} R^a}{\mathrm{d} \hat{t}} &= M^{ab}P_b,\label{eq16a}\\
        \frac{\mathrm{d} P_a}{\mathrm{d} \hat{t}} &= -\nabla_a \hat{W}.\label{eq16b}
    \end{eqnarray}
\endnumparts
Combining (\ref{eq16a}) and (\ref{eq16b}), the dynamics equation becomes:
\begin{eqnarray}
    m\frac{\mathrm{d}^2 R^a}{\mathrm{d} \hat{t}^2} = -m M^{ab}\nabla_b \hat{W} \equiv F^a,\label{eq17}
\end{eqnarray}
where the effective scalar mass $m=2/M^a{}_a$. Note that (\ref{eq17}) could
also be derived by multiplying $m M^{ab}$ at both side of (\ref{eq12}).
(\ref{eq17}) shows the dynamics of an anisotropic mass $m_{ab}$ under the
potential $\hat{W}$, which is equivalent to the dynamics of an isotropic mass
$m$ subject to a non-conservative force $F^a$.

By decomposing $M^{ab}$ as
\begin{eqnarray}
    M^{ab} = \frac{1}{m}\delta^{ab} + \Omega^{ab},\label{eq18}
\end{eqnarray}
the total force $F^a$ is divided into the conservative part $f^a$ and
the non-conservative part $\bar{f}^a$:
\begin{eqnarray}
    F^a &= f^a + \bar{f}^a\nonumber\\
    &= - \nabla^a \hat{W} - m\Omega^{ab} \nabla_b \hat{W}.\label{eq19}
\end{eqnarray}
One may check that in general the curl of $F^a$ is nonzero:
\begin{eqnarray}
    (\mathrm{d} F)_{\mu\nu} &= \partial_{\mu} \bar{f}_{\nu} - \partial_{\nu}
    \bar{f}_{\mu}\nonumber\\
    &= - \frac{2m_2}{2m_1 + m_2}\frac{\partial^2 \hat{W}}{\partial X_1 \partial Y_1}.\label{eq22}
\end{eqnarray}

Now, we analysis the work done by $F^a$. 
Following the discussion in the main text, the total kinetic energy could be
separated into the isotropic part $T_\mathrm{I}$ and the anisotropic part $T_\mathrm{A}$ as:
\begin{eqnarray}
  T &= T_\mathrm{I} - T_\mathrm{A} \nonumber\\
  &= \frac12 m \left(u^2 + v^2\right) - \frac12 \left[\left(m- m_1 - m_2\right)u^2 + \left(m - m_1\right)v^2\right].\label{eq20}
\end{eqnarray}

In a cyclic motion of $R^a$ along a closed orbit, the work done by the
conservative force $f^a$ is equal to the change in the total kinetic energy,
denoted as $\Delta T$. Due to the energy conservation, $\Delta T = -\Delta
\hat{W} = 0$. Therefore, $\Delta T_\mathrm{A}$ is equal to $\Delta T_\mathrm{I}$. According to
(\ref{eq20}), we obtain the following relations:
\begin{eqnarray}
    \Delta T_\mathrm{I} = \frac12 m \Delta\left(u^2+v^2\right),\quad
    \Delta T_\mathrm{A} = \frac{m-m_1}{m}\Delta T_\mathrm{I} - \frac12 m_2 \Delta (u^2).
\end{eqnarray}
Consequently,
\begin{eqnarray}
    \frac{m_1}{m}\Delta T_\mathrm{I} = - \frac12 m_2 \Delta (u^2).\label{eq26}
\end{eqnarray}
(\ref{eq26}) reveals that in the
cyclic motion of $R^a$, the net work extracted by the non-conservative force
$F^a$, which is $\Delta T_\mathrm{I}$, is originated from the
reduction of the kinetic energy of $m_2$ only.

\section{Conclusions}

In summary, from the perspective of geometry, we model a $d$-dimensional
continuum elastic body as a Riemannian manifold embedded in the
$(d+1)$-dimensional Euclidean space, and construct a Hamiltonian framework for
the elastic body of finite strain. It is shown that the antisymmetry of the
elastic modulus tensor is originated from the anisotropic mass tensor in the
kinetic term. The resulting odd elastic modulus exhibits inherent anisotropy. We
also derive the non-conservative stress and the associated nonlinear
constitutive relation, where the nonlinearity is caused by the intrinsic
geometry of the deformed elastic body.  The Hamiltonian formalism constructed in
this work for characterizing odd elasticity also allows one to explore the
physics of odd elasticity in dynamical regime. We shall finally point out
that our findings also raise some questions to be explored: Can the two features
of antisymmetry and anisotropy of the elastic modulus be separated? Are there
any other mechanisms to generate the characteristic antisymmetry of the
odd-elastic modulus in the frame of Hamiltonian formalism?

\ack
This work was supported by the National Natural Science Foundation of China
(Grants No. BC4190050).

\appendix


\section{Expressions for the induced metric and strain}\label{AA}

In this section, we derive the expressions for the induced metric and strain.
Note that we employ the abstract index notation in Latin letters to represent a
tensor; the tensor components are labeled by Greek letters.  Abstract indices
indicate the type of a tensor, with repeated indices in pairs denoting tensor
contraction. The convention of Einstein summation is applied to repeated Greek
indices.

First of all, for any local coordinates system $\{s^1, \dots, s^d\}$ on $\mathcal{B}$, the
associated coordinates base vector on its embedding
\begin{eqnarray}
    \left(\frac{\partial}{\partial s^\mu}\right)^a = \frac{\partial \bar{X}^\nu}{\partial s^\mu}\left(\frac{\partial}{\partial \bar{X}^\nu}\right)^a ,
\end{eqnarray}
where $\bar{X}^\nu$ is the Cartesian coordinates of the embedding
of deformed configuration $\Sigma$ in the Euclidean space $\mathcal{E}$.

The isometric embedding of $\Sigma$ in the Euclidean space $\mathcal{E}$
indicates that 
\begin{eqnarray}
\bar{g}_{ab} = \delta_{ab} - N_a N_b,
\end{eqnarray}
where $N_a$ is the unit normal covector on $\Sigma$. Therefore, the component of
the metric tensor $\bar{g}_{ab}$ is
\begin{eqnarray}
   \bar{g}_{\mu\nu}&=\bar{g}_{ab}\left(\frac{\partial}{\partial s^\mu}\right)^a\left(\frac{\partial}{\partial s^\nu}\right)^b\nonumber \\
   &=\left(\delta_{ab}-N_aN_b\right)\left(\frac{\partial}{\partial s^\mu}\right)^a\left(\frac{\partial}{\partial s^\nu}\right)^b\nonumber \\
   &=\delta_{ab}\left(\frac{\partial}{\partial s^\mu}\right)^a\left(\frac{\partial}{\partial s^\nu}\right)^b\nonumber \\
   &=\delta_{\tau\rho}\frac{\partial \bar{X}^\tau}{\partial s^\mu}\frac{\partial \bar{X}^\rho}{\partial s^\nu} \nonumber \\
   &= \frac{\partial \bar{X}_\rho}{\partial s^\mu}\frac{\partial \bar{X}^\rho}{\partial s^\nu}.\label{barg_munu}
\end{eqnarray}
Here $\delta_{ab}$ is the Euclid metric of $\mathcal{E}$, and $\delta_{\tau\rho}$
is Kronecker delta which is the components of $\delta_{ab}$ in Cartesian
coordinates. In the derivation of (\ref{barg_munu}),
we use the fact that
$\left(\partial/\partial{s^\mu}\right)^aN_a = 0$. One has the similar result for $g_{\mu\nu}$
as
\begin{eqnarray}
    g_{\mu\nu}=\frac{\partial X_{0\rho}}{\partial s^\mu}\frac{\partial X_0^\rho}{\partial s^\nu},\label{g_munu}
\end{eqnarray}
where $X_0^\rho$ is the Cartesian coordinates of the embedding of reference
configuration.

By substituting (\ref{barg_munu}) and (\ref{g_munu}) into the following
expression of $u_{ab}$
\begin{eqnarray}
    u_{ab}=\frac{1}{2}\left(\bar{g}_{ab}-g_{ab}\right),\ \label{a_def_strain}
\end{eqnarray}
we obtain
\begin{eqnarray}
  u_{ab} &=
    \frac{1}{2}\left(\frac{\partial \bar{X}_{\rho}}{\partial s^{\mu}}\frac{\partial \bar{X}^{\rho}}{\partial s^{\nu}}-\frac{\partial X_{0\rho}}{\partial s^{\mu}}\frac{\partial X_{0}^{\rho}}{\partial s^{\nu}}\right)\left(\mathrm{d}
    s^{\mu}\right)_a\left(\mathrm{d} s^{\nu}\right)_b\nonumber\\
    &= \frac{1}{2}\left(\nabla_a \bar{X}_\rho \nabla_b \bar{X}^\rho - \nabla_a X_{0\rho} \nabla_b X_{0}^{\rho}\right) ,\label{components_uab}
\end{eqnarray}
where $\nabla_b$ is the covariant derivative operator on $\Sigma$. $\nabla_b$ is
metric compatible, i.e., $\nabla_b\bar{g}_{ac}=0$. 
(\ref{components_uab}) shows that $u_{ab}$ could be regarded as a local
function of $\nabla_a \bar{X}^\rho$.

\section{Variational derivatives of the Hamiltonian}\label{AB}

In this section, we present the variational derivatives of the Hamiltonian
defined on the manifold $\mathcal{B}$, which is recorded here
\begin{eqnarray}
    H &= T + W\nonumber\\
    &= \frac{1}{2}\int\mathrm{d}^{d}s\rho_0\sqrt{g}\mathscr{M}^{ab} P_a P_b + \int\mathrm{d}^{d}s\rho_0\sqrt{g}\mathscr{W} . \label{kinetic_energy}
\end{eqnarray}

The variational derivative of the Hamiltonian with respect to the
generalized momentum density field $P_a$ is
\begin{eqnarray}
   \frac{\delta H}{\delta P_a} = \frac{\delta T}{\delta P_a} = \rho_0\sqrt{g}\mathscr{M}^{ab} P_b .\label{fdP}
\end{eqnarray}

To obtain the variational derivative of the Hamiltonian with respect to $U^a$,
we first calculate
\begin{eqnarray}
    \delta W &= \int\mathrm{d}^{d}s\rho_0\sqrt{g}\delta\mathscr{W}\nonumber\\
    &=\int\mathrm{d}^{d}s\rho\sqrt{\bar{g}} \frac{\partial \mathscr{W}}{\partial u_{ab}}\delta u_{ab}\nonumber\\
    &=\int\mathrm{d}^{d}s\sqrt{\bar{g}} \sigma^{ab} \nabla_a \bar{X}_{\rho} \nabla_b \delta\bar{X}^{\rho}\nonumber\\
    &=\int\mathrm{d}^{d}s\sqrt{\bar{g}}\nabla_b\left(\sigma^{ab} \nabla_a \bar{X}_{\rho} \delta\bar{X}^{\rho}\right) - \int\mathrm{d}^{d}s\sqrt{\bar{g}}\nabla_b\left(\sigma^{ab} \nabla_a \bar{X}_{\rho}\right) \delta\bar{X}^{\rho}\nonumber\\
    &=- \int\mathrm{d}^{d}s\sqrt{\bar{g}}\nabla_b\left(\sigma^{ab} \nabla_a \bar{X}_{\rho}\right) \delta\bar{X}^{\rho},\label{var_free_energy}
\end{eqnarray}
Note that from the first line to the second line in
(\ref{var_free_energy}), the
domain of integration is changed from the reference configuration to $\Sigma$. 
In the last
equality, we drop the
divergence term to the boundary term by utilizing Gauss's theorem; only the
energy variation in the interior of the body is considered.

Now, we calculate the variational derivative of the Hamiltonian with respect to
the displacement field $U^a$:
\begin{eqnarray}
   \frac{\delta H}{\delta U^a} &= \frac{\delta W}{\delta U^a} \nonumber\\
    &= \frac{\delta W}{\delta \bar{X}^\mu}\left(\mathrm{d} \bar{X}^{\mu}\right)_a\nonumber\\ 
    &=-\sqrt{\bar{g}}\left(\mathrm{d} \bar{X}^{\mu}\right)_a\nabla_c\left(\sigma^{bc}\nabla_b \bar{X}_{\mu} \right)\nonumber\\
    &=-\sqrt{\bar{g}}\left(\mathrm{d} \bar{X}^{\mu}\right)_a \bar{g}_c{}^d \partial_d\left[\sigma^{bc} \bar{g}_b{}^e\left(\mathrm{d}\bar{X}_{\mu}\right)_e\right]\nonumber\\
    &=-\sqrt{\bar{g}} \bar{g}_c{}^d \partial_d\left[\sigma^{bc} \bar{g}_b{}^e \left(\mathrm{d} \bar{X}^{\mu}\right)_a\left(\mathrm{d}\bar{X}_{\mu}\right)_e\right]\nonumber\\
    &=-\sqrt{\bar{g}} \bar{g}_c{}^d \partial_d\left(\sigma^{bc} \bar{g}_b{}^e \delta_{ae}\right)\nonumber\\
    &=-\sqrt{\bar{g}} \bar{g}_c{}^d \partial_d\left(\bar{g}_{ab}\sigma^{bc}\right) \nonumber\\ 
    &\equiv -\sqrt{\bar{g}} f_a. \label{fdU}
\end{eqnarray}
$\partial_a$ is the derivative operator in the Cartesian coordinates on
$\mathcal{E}$; $\partial_{a}\left(\mathrm{d} \bar{X}_{\mu}\right)_b=0$. $(\mathrm{d})_a$ is the
exterior derivative operator on $\mathcal{E}$.
$\bar{g}_a{}^b$ is a projection operator: $\bar{g}_a{}^b =
\bar{g}_{ac}\delta^{cb}=\delta_a{}^b-N_aN^b$.

Here, it is of interest to point out that in general $f_a$ is not tangent to
$\Sigma$, which is shown below 
\begin{eqnarray}
    f_a&= \bar{g}_c{}^d \partial_d\left(\bar{g}_{ab}\sigma^{bc}\right)
    \nonumber\\
    &= \bar{g}_c{}^d \partial_d\left(\sigma^{bc} \bar{g}_b{}^e \delta_{ae}\right)\nonumber\\
    &=\delta_{ae}\bar{g}_c{}^d \partial_d\left(\sigma^{bc} \bar{g}_b{}^e\right)\nonumber\\
    &=\delta_{ae}\left(\bar{g}_b{}^e \bar{g}_c{}^d \partial_d\sigma^{bc} + \sigma^{bc} \bar{g}_c{}^d \partial_d \bar{g}_b{}^e\right)\nonumber\\
    &=\delta_{ae}\left(\nabla_c\sigma^{ec}-\sigma^{bd}N^e\partial_dN_b\right)\nonumber\\
    &=\bar{g}_{ab}\nabla_c\sigma^{bc}-N_a\sigma^{bc}K_{bc},\label{fdU2}
 \end{eqnarray}
where $K_{bc}$ is the extrinsic curvature of $\Sigma$. In the derivation, we use
the definition of extrinsic curvature
$K_{bc}=\bar{g}_b{}^d\bar{g}_c{}^e\partial_dN_e$ and the
fact that $\sigma^{ab}N_b=0$.
The first and second terms in (\ref{fdU2}) are the tangent and normal
components of $f_a$~\cite{Guven2002,Guven2004a,Guven_2004}. 
For example, in the case of $\phi_2$ in figure~\ref{fig1} of the main text, the normal
force supporting $\Sigma$ on the sphere is $(e_r)^a\sigma^b{}_b/R$, where $R$
is the radius of the sphere and $(e_r)^a$ is the unit normal vector
perpendicular to the spherical surface.

\begin{figure}[!ht]
    \begin{subfigure}[t]{0.4\textwidth}
           \includegraphics[width=\textwidth]{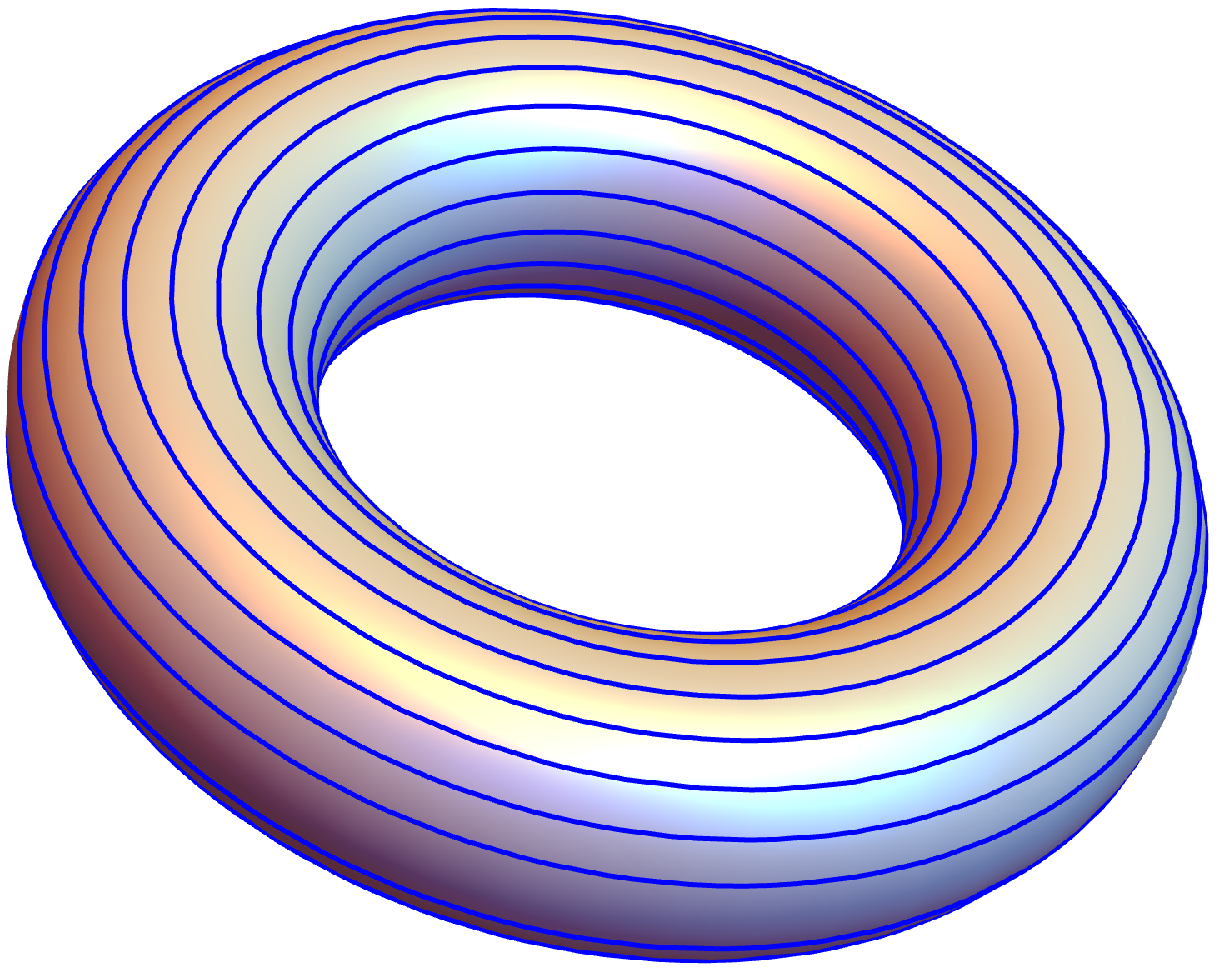}
            \caption{Torus}
            \label{fig:1a}
    \end{subfigure}
    \quad \quad \quad \quad \quad \quad
    \begin{subfigure}[t]{0.4\textwidth}
            \includegraphics[width=\textwidth]{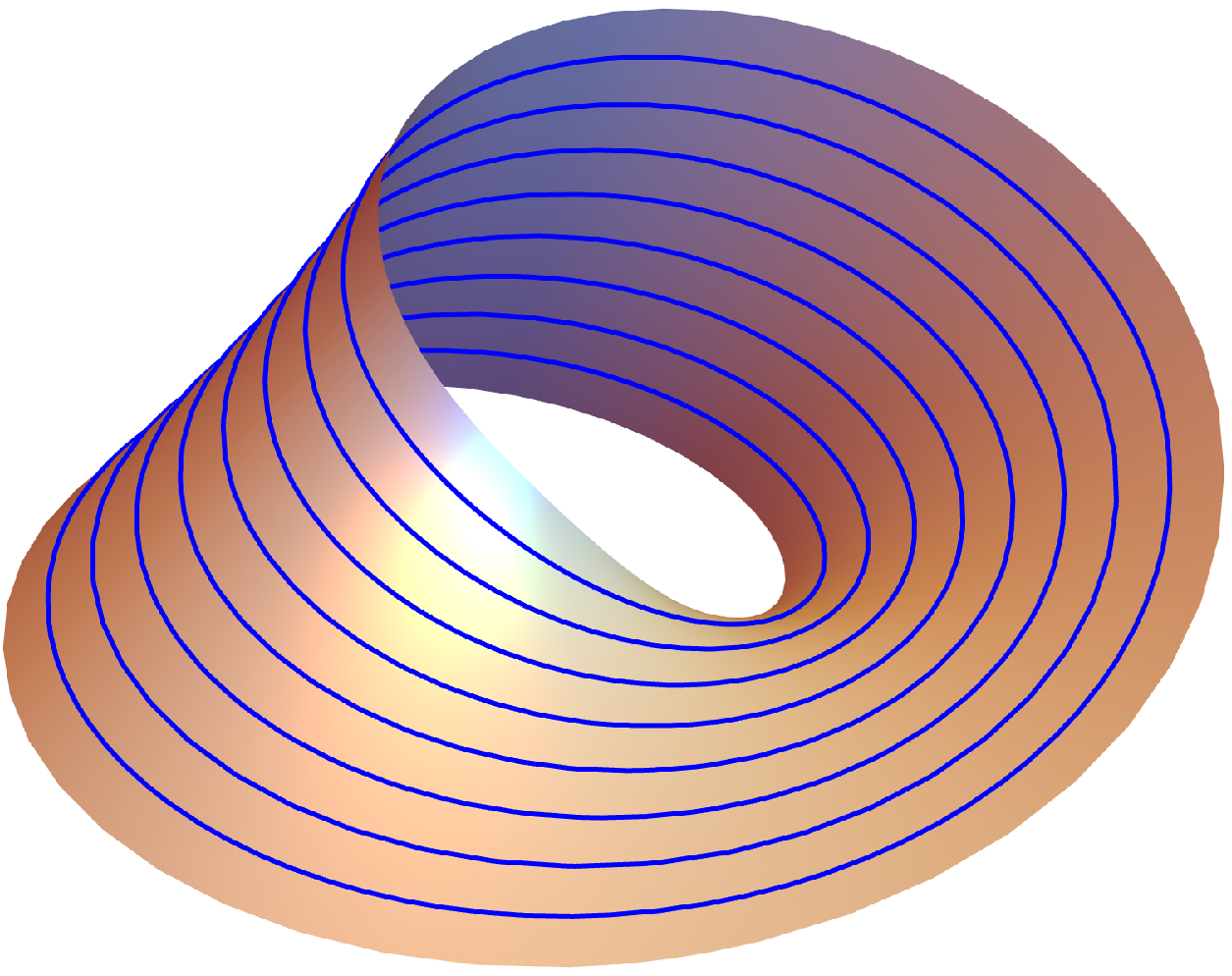}
            \caption{M\"{o}bius strip}
            \label{fig:1b}
    \end{subfigure}
    \caption{Illustrations of 2D torus and M\"{o}bius strip. The tangent
    vectors of the lines constitute the Killing vector field on these two
    surface.  
    \label{figS1}}
\end{figure}

\section{The isometric symmetry of elastic modulus tensor}\label{AC}

In the main text, we present the following expression for the rotationally
symmetric elastic modulus tensor in an axisymmetric continuum elastic body:
\begin{eqnarray}
    C^{abcd}=\lambda g^{ab}g^{cd}+2\mu g^{a(c}g^{d)b},\label{A_C}
\end{eqnarray}
where $\lambda$ and $\mu$ are Lam\'e coefficients. In this section, we present
the derivation for (\ref{A_C}).

Consider an axisymmetric continuum elastic body. Its rotation as a whole
belongs to a special isometric mapping that could be generated by a Killing
vector field $v^a$. By the definition of the Killing vector
field, the displacement of the points from
$s^{\mu}$ to $s^{\mu}+v^{\mu} \epsilon$ leaves
the distance relationships unchanged. In other words, the displacement of
$v^{\mu} \epsilon$ defines an isometric mapping.
In figure~\ref{figS1}(a), we present a 2D torus as an example of the axisymmetric
elastic body. The tangent vectors of the toroidal lines constitute the Killing
vector field on the torus. The elastic modulus tensor as associated with
infinitesimal volume element possesses the rotational symmetry in the sense
that it is invariant along the toroidal lines.

The rotational symmetry of the elastic modulus tensor $C^{abcd}$ is expressed
by the zero Lie derivative of $C^{abcd}$ along the direction as specified by the
Killing vector on the reference configuration~\cite{lang2012}:  
\begin{eqnarray}
    \mathscr{L}_v C^{abcd} &\equiv& v^e \nabla_e C^{abcd} - C^{ebcd} \nabla_e v^a - C^{aecd} \nabla_e v^b\nonumber\\
    &&- C^{abed} \nabla_e v^c - C^{abce} \nabla_e v^d\nonumber\\
    &=& 0, \label{LieC}
\end{eqnarray}
where $v^a$ is a Killing vector field satisfying $\nabla_{(a} v_{b)} = 0$, and
$\nabla_a$ is the covariant derivative operator on the reference configuration.
$C^{abcd}$ as constructed by the tensor product of $g^{ab}$-tensors satisfies
(\ref{LieC}); note that $\mathscr{L}_v g^{ab}=0$. Specifically, there are
three kinds of index permutations of the tensor product:
\begin{eqnarray}
    C^{abcd}=\lambda g^{ab}g^{cd}+\mu_1 g^{ac}g^{bd}+\mu_2 g^{ad}g^{bc},\label{general_C}
\end{eqnarray}
where $\lambda$, $\mu_1$ and $\mu_2$ are constant scalars on $\mathcal{B}$.

Due to the extra requirements for the conservation of angular momentum and the conservative
nature of stress~\cite{Scheibner2020}, 
\begin{eqnarray}
    C^{abcd}=C^{cdab}=C^{bacd}\ .\label{index_symmetry}
\end{eqnarray} 
The first equality is automatically satisfied for (\ref{general_C}).
The second equality is satisfied if $\mu_1=\mu_2=\mu$. The final expression for
$C^{abcd}$ that satisfies both (\ref{general_C}) and (\ref{index_symmetry})
is
\begin{eqnarray}
    C^{abcd}=\lambda g^{ab}g^{cd}+2\mu g^{a(c}g^{d)b} .\label{Cabcd}
\end{eqnarray}

For Euclidean reference configuration, the components of $C^{abcd}$ in
Cartesian coordinates system are
\begin{eqnarray}
    C^{\mu\nu\eta\gamma}=\lambda \delta^{\mu\nu}\delta^{\eta\gamma}+2\mu \delta^{\mu(\eta}\delta^{\gamma)\nu} .\label{components_C}
\end{eqnarray}
By making use of the linear constitutive relation, the elastic potential can be expressed as
\begin{eqnarray}
    W = \int\mathrm{d}^{d}s\rho_0\left[\lambda
    \left(u_{\eta\eta}\right)^2+2\mu\left(u_{\eta\gamma}\right)^2\right], \label{ON_free_energy}
\end{eqnarray}
where $\mathrm{d}^{d}s$ is the volume element of Cartesian coordinates on the reference
configuration.  In comparison with the linear elasticity theory, the parameters
$\lambda$ and $\mu$ are recognized as the Lam\'e coefficients~\cite{Landau1960}.

Note that (\ref{LieC}) could be used to describe the general case that the
elastic modulus tensor is invariant along the lines of the Killing vectors,
i.e., the isometric symmetry of the elastic modulus tensor. For example, on the
M\"{o}bius strip in figure (\ref{fig:1b}), the tangent vectors of the lines
constitute the Killing vector field; the isometric mapping of the M\"{o}bius
strip is generated by this Killing vector field. $C^{abcd}$ that satisfies
(\ref{LieC}) is invariant along these lines. This is the generalization of
the rotational symmetry of the elastic modulus tensor of axisymmetric continuum
elastic body as discussed in this section.

\section{The symmetry of $\hat{C}^{abcd}$ and $\hat{D}^{abcdef}$}\label{AD}
In the main text, we assert that the odd elastic modulus $\hat{C}^{abcd}$ and
$\hat{D}^{abcdef}$ possess inherent anisotropy, which is different from
traditional odd elastic modulus. Following the discussion in
\ref{AC} about the symmetry of elastic modulus tensor, we can now
further supplement the symmetry of $\hat{C}^{abcd}$ and $\hat{D}^{abcdef}$ in
details.

According to the definition in \ref{AC}, the sufficient and
necessary condition for odd elastic modulus to be isometric is
satisfied if $\mathscr{L}_v\hat{C}^{abcd} = 0$ and
$\mathscr{L}_v\hat{D}^{abcdef} = 0$. Specifically, this requires $C^{abcd}$ to
satisfy the following equation:
\begin{eqnarray}
    \rho\Omega^{ae} g_{ef} \mathscr{L}_v C^{fbcd} + g_{ef} C^{fbcd} \mathscr{L}_v(\rho\Omega^{ae}) = 0,\label{iso_Cbar}
\end{eqnarray}
where the isometric symmetry of $\hat{C}^{abcd}$ is satisfied; and:
\begin{eqnarray}
    \rho\Omega^{ac} \mathscr{L}_vC^{dbef} + C^{dbef}\mathscr{L}_v(\rho\Omega^{ac}) = 0,\label{iso_Dbar}
\end{eqnarray}
where the isometric symmetry of $\hat{D}^{abcdef}$ is satisfied.
Substituting (\ref{iso_Dbar}) into (\ref{iso_Cbar}) and eliminating
the Lie derivative of $\rho\Omega^{ab}$, we obtain an identity.
Therefore, (\ref{iso_Cbar}) and (\ref{iso_Dbar}) are equivalent,
meaning the symmetries of $\hat{C}^{abcd}$ and $\hat{D}^{abcdef}$ are
consistent. In the following, we perform a detailed calculation of the
symmetry of $\hat{C}^{abcd}$. The final result shows that, due to the
coupling with $\rho\Omega^{ab}$, $\hat{C}^{abcd}$ is necessarily anisotropic.
    
Generally, solving for Killing vector fields on a manifold is challenging.
However, for simple cases, the corresponding Killing vector fields can be
directly written using known properties of manifold symmetries, and the
isometric condition for odd elastic modulus can be derived using
(\ref{LieC}) or (\ref{iso_Cbar}).

First, let's consider the case where the reference configuration is a 2D
Euclidean plane ($d=2,\ g_{ab}=\delta_{ab}$).
    
The 2D Euclidean plane has 3 independent symmetries: translations along the
$x$ and $y$ axes, and rotation about the origin. The corresponding 3 Killing
vector fields are: $e_1{}^a \equiv (\partial/\partial{x})^a$,
$e_2{}^a \equiv (\partial/\partial{y})^a$,
$(\partial/\partial{\varphi})^a \equiv -y(\partial/\partial{x})^a+x(\partial/\partial{y})^a$,
where $\varphi$ is the polar angle. Below, we denote the Lie derivative
operators along these three Killing vector fields as
$\mathscr{L}_x$, $\mathscr{L}_y$, $\mathscr{L}_{\varphi}$.
In this context, the translational and rotational symmetry of the 2D
Euclidean plane correspond to the homogeneous and isotropy of the elastic
modulus, respectively.
    
The analysis is divided into two steps: first, to determine the general form of the odd elastic modulus $\hat{C}^{abcd}$ that satisfies the homogeneous
isotropic condition, and then to find the $C^{abcd}$ that makes
$\hat{C}^{abcd}$ satisfy this condition.
    
We begin with (\ref{LieC}) to give the homogeneous isotropic condition
for $\hat{C}^{abcd}$, that is, $\mathscr{L}_v \hat{C}^{abcd}=0$. In Cartesian
coordinates, $\hat{C}^{abcd}$ can be expanded as:
\begin{eqnarray}
    \hat{C}^{abcd} = \hat{C}^{\mu\nu\eta\gamma} e_\mu{}^a e_\nu{}^b e_\eta{}^c e_\gamma{}^d.\label{cartesian_hatC}
\end{eqnarray}
If and only if $\hat{C}^{\mu\nu\eta\gamma}$ are position-independent constant
components, the following equations hold:
\begin{eqnarray}
    \mathscr{L}_x \hat{C}^{abcd} = \frac{\partial \hat{C}^{\mu\nu\eta\gamma}}{\partial x} e_\mu{}^a e_\nu{}^b e_\eta{}^c e_\gamma{}^d = 0,\\
    \mathscr{L}_y \hat{C}^{abcd} = \frac{\partial \hat{C}^{\mu\nu\eta\gamma}}{\partial y} e_\mu{}^a e_\nu{}^b e_\eta{}^c e_\gamma{}^d = 0.
\end{eqnarray}
In this case, $\hat{C}^{abcd}$ automatically satisfies the translational
symmetry (homogeneous), and only the rotational symmetry (isotropy) condition
$\mathscr{L}_{\varphi}\hat{C}^{abcd}=0$ needs to be considered. This
condition is equivalent to the polar coordinate components of
$\hat{C}^{abcd}$ being independent of $\varphi$.

In the following, we derive for the rotational symmetry
condition of $\hat{C}^{abcd}$ in 2D.

The polar coordinate components $\hat{C}_\varphi{}^{\mu\nu\eta\gamma}$
of $\hat{C}^{abcd}$ are obtained as follows:
\begin{eqnarray}
    \hat{C}_\varphi{}^{\mu\nu\eta\gamma} &= J_\varphi{}^{\mu}{}_{\alpha} J_\varphi{}^{\nu}{}_{\beta} J_\varphi{}^{\eta}{}_{\rho} J_\varphi{}^{\gamma}{}_{\sigma}\hat{C}^{\alpha\beta\rho\sigma}\nonumber\\
    &= J_\varphi{}^{\mu}{}_{\alpha} J_\varphi{}^{\nu}{}_{\beta} J_\varphi{}^{\eta}{}_{\rho} J_\varphi{}^{\gamma}{}_{\sigma} e^\alpha{}_a e^\beta{}_b e^\rho{}_c e^\sigma{}_d \mathfrak{e}_\tau{}^{ab} \mathfrak{e}_\theta{}^{cd} \hat{C}^{\tau\theta},
\end{eqnarray}
where the Jacobian matrix
$J_\varphi{}^{\mu}{}_{\nu} \equiv \partial \left(\varphi,r\right)/\partial \left(x,y\right)$ transforms Euclidean coordinates to polar coordinates.
The sufficient and necessary condition for $\hat{C}^{abcd}$ to have
rotational symmetry is:
\begin{eqnarray}
   \frac{\partial \hat{C}_\varphi{}^{\mu\nu\eta\gamma}}{\partial \varphi} = e^\alpha{}_a e^\beta{}_b e^\rho{}_c e^\sigma{}_d \mathfrak{e}_\tau{}^{ab} \mathfrak{e}_\theta{}^{cd} \hat{C}^{\tau\theta} \frac{\partial}{\partial \varphi}\left(J_\varphi{}^{\mu}{}_{\alpha} J_\varphi{}^{\nu}{}_{\beta} J_\varphi{}^{\eta}{}_{\rho} J_\varphi{}^{\gamma}{}_{\sigma}\right) = 0
\end{eqnarray}
In deriving the above equation, we use the condition that
$\hat{C}^{\mu\nu\eta\gamma}$, and thus $\hat{C}^{\tau\theta}$,
is a constant matrix. Solving this matrix equation for
$\hat{C}^{\tau\theta}$, we get:
\begin{eqnarray}
    \fl \hat{C}^{11}=\hat{C}^{11}, \quad \hat{C}^{12}=\hat{C}^{12}, \quad \hat{C}^{21}=\hat{C}^{21},\quad \hat{C}^{22}=\hat{C}^{22}, \quad \hat{C}^{43}=-\hat{C}^{34},\quad \hat{C}^{33}=\hat{C}^{44},
\end{eqnarray}
with all other components being zero. When expressed in matrix form, it
becomes:
\begin{eqnarray}
    \hat{C}^{\tau\theta }=
    \begin{bmatrix}
        \hat{C}^{11} & \hat{C}^{12} & 0 & 0 \\
        \hat{C}^{21} & \hat{C}^{22} & 0 & 0 \\
        0 & 0 & \hat{C}^{33} & \hat{C}^{34} \\
        0 & 0 & -\hat{C}^{34} & \hat{C}^{33}
    \end{bmatrix},\label{general_hatC}
\end{eqnarray}
(\ref{general_hatC}) represents the isotropic condition for
$\hat{C}^{abcd}$, with 6 independent stiffness coefficients, consistent with
the results in the literature~\cite{Scheibner2020}.

Next, we construct $C^{abcd}$ that makes $\hat{C}^{\mu\nu}$ satisfying
(\ref{general_hatC}).

Since $\hat{C}^{abcd}=\rho\Omega^{ae} \delta_{ef} C^{fbcd}$, we have:
\begin{eqnarray}
    \hat{C}^{\mu\nu} = \rho\Omega^{ae} \delta_{ef} C^{fbcd}\mathfrak{e}^\mu{}_{ab}\mathfrak{e}^\nu{}_{cd}. \label{condition_hatC_isotropy}
\end{eqnarray}
Let $C^{abcd}$ be expanded in the frame $\{\mathfrak{e}^\mu{}_{ab}\}$ as:
\begin{eqnarray}
    C^{abcd} = C^{\mu\nu}\mathfrak{e}_\mu{}^{ab}\mathfrak{e}_\nu{}^{cd},\label{cartesian_C}
\end{eqnarray}
where $C^{\mu\nu}$ has the index symmetry, i.e., $C^{\mu\nu}=C^{\nu\mu}$.
Substituting (\ref{tensor_frame}), (\ref{anisotropic_mass_component}) and(
\ref{cartesian_C}) into (\ref{condition_hatC_isotropy}), we obtain the
following matrix equation for $C^{\mu\nu}$:
\begin{eqnarray}
    \hat{C}^{\mu\nu} = C^{\eta\gamma}\left(\kappa\mathfrak{e}_3{}^{ae} + \zeta\mathfrak{e}_4{}^{ae}\right)\mathfrak{e}^1{}_{ef}\mathfrak{e}_\eta{}^{fb}\mathfrak{e}_\gamma{}^{cd}\mathfrak{e}^\mu{}_{ab}\mathfrak{e}^\nu{}_{cd}.\label{2D_isocondition}
\end{eqnarray}
The condition for the solution of (\ref{2D_isocondition}) is that
$\rho\Omega^{ab}$ is non-degenerate, and the upper-left submatrix of matrix
$\hat{C}^{\tau\theta}$ is equal to the lower-right submatrix, namely:
\begin{eqnarray}
    \hat{C}^{\tau\theta} =
    \begin{bmatrix}
        \hat{C}^{33} & \hat{C}^{34} & 0 & 0 \\
        -\hat{C}^{34} & \hat{C}^{33} & 0 & 0 \\
        0 & 0 & \hat{C}^{33} & \hat{C}^{34} \\
        0 & 0 & -\hat{C}^{34} & \hat{C}^{33}
    \end{bmatrix}.\label{iso_odd}
\end{eqnarray}
This represents a specific odd elastic modulus with only 2 independent
stiffness coefficients. The solution to (\ref{2D_isocondition}) is then:
\begin{eqnarray}
    \fl C^{\mu\nu} =\frac{1}{4\left(\zeta^2 + \kappa^2\right)}
    \begin{bmatrix}
        0 & 0 & \kappa \hat{C}^{33} - \zeta \hat{C}^{34} & \zeta \hat{C}^{33} + \kappa \hat{C}^{34} \\
        0 & 0 & \zeta \hat{C}^{33} + \kappa \hat{C}^{34} & \zeta \hat{C}^{34} - \kappa \hat{C}^{33} \\
        \kappa \hat{C}^{33} - \zeta \hat{C}^{34} & \zeta \hat{C}^{33} + \kappa \hat{C}^{34} & 0 & 0 \\
    \zeta \hat{C}^{33} + \kappa \hat{C}^{34} & \zeta \hat{C}^{34} - \kappa \hat{C}^{33} & 0 & 0
    \end{bmatrix}.\label{iso_odd_condition}
\end{eqnarray}
Thus, by constructing a $C^{abcd}$ on the elastic body that satisfies
($\ref{iso_odd_condition}$) (which is evidently an anisotropic elastic
modulus), one can obtain the homogeneous isotropic odd elastic modulus given
by ($\ref{iso_odd}$).

According to our assumption, the rotation of the elastic body does not change
its intrinsic geometry, and thus does not induce stress. Therefore,
$C^{abcd}$ will also satisfy $C^{abcd}=C^{abdc}$, or $C^{\mu 2}=0$. At this
point, (\ref{2D_isocondition}) only has the trivial solution
$C^{\mu\nu}=\hat{C}^{\mu\nu}=0$. This implies that for a reference
configuration of a 2D Euclidean plane, the odd elastic modulus
$\hat{C}^{abcd}$ and $\hat{D}^{abcdef}$ must be anisotropic due to the
coupling with $\rho\Omega^{ab}$. It is known that any sufficiently small
neighborhood of a point on a Riemannian manifold can be considered as a
Euclidean space of the same dimension. Therefore, for any case of a 2D
elastic body reference configuration, the odd elastic modulus
$\hat{C}^{abcd}$ and $\hat{D}^{abcdef}$ are anisotropic tensor fields.

Finally, some additional remarks on the assumption of $C^{\mu 2}=0$ are
warranted. This assumption is known as the ``objectivity'' of the elastic
body~\cite{Landau1960,Scheibner2020}, predicated on the premise that the
interaction forces between the microscopic units of the elastic body depend
only on their relative distance and not on the orientation. When a substrate
exists, this condition may no longer hold~\cite{Braverman2021,Nelson1979}.
Thus, a more accurate statement is that, for any 2D elastic body
possessing objectivity, the odd elastic modulus $\hat{C}^{abcd}$ and
$\hat{D}^{abcdef}$ are anisotropic tensor fields.

Next, we apply the above discussion to the case where the reference
configuration is a 3D Euclidean space ($d=3,\ g_{ab}=\delta_{ab}$).

A 3D Euclidean space has 6 independent Killing vector fields, corresponding
to translational symmetry along the $x$, $y$, and $z$ axes:
$e_1{}^a \equiv (\partial/\partial{x})^a$, $e_2{}^a \equiv (\partial/\partial{y})^a$,
$e_3{}^a \equiv (\partial/\partial{z})^a$; and rotational symmetry around the $z$, $x$,
and $y$ axes:
$(\partial/\partial{\vartheta})^a \equiv -y(\partial/\partial{x})^a + x(\partial/\partial{y})^a$,
$(\partial/\partial{\varrho})^a \equiv -z(\partial/\partial{y})^a + y(\partial/\partial{z})^a$,
$(\partial/\partial{\varsigma})^a \equiv -x(\partial/\partial{z})^a + z(\partial/\partial{x})^a$.
Assuming that the Cartesian components $\hat{C}^{\mu\nu\eta\gamma}$ of
$\hat{C}^{abcd}$ are still constant matrices, the homogeneous is
automatically satisfied, and only its isotropic properties, i.e., rotational
symmetry, need to be considered. We will next calculate the Lie derivatives
$\mathscr{L}_\vartheta$, $\mathscr{L}_\varrho$, and $\mathscr{L}_\varsigma$
corresponding to the rotational symmetry around the $z$, $x$, and $y$ axes
for $\hat{C}^{abcd}$.

To facilitate the consideration of rotational symmetry around the $z$, $x$,
and $y$ axes, three sets of cylindrical coordinates are introduced as follows:
\begin{eqnarray}
    \left\{
    \begin{array}{rcl}
        x = r \cos{\vartheta},\quad y = r \sin{\vartheta},\quad z = z; \\
        x = x,\quad y = r \cos{\varrho},\quad z = r \sin{\varrho}; \\
        x = r \sin{\varsigma},\quad y = y,\quad z = r \cos{\varsigma}.
    \end{array}\right.
\end{eqnarray}
The components of $\hat{C}^{abcd}$ in the three sets of cylindrical
coordinates, $\hat{C}_\lambda{}^{\mu\nu\eta\gamma}$
($\lambda=\vartheta,\varrho,\varsigma$), can be obtained as follows:
\begin{eqnarray}
    \hat{C}_\lambda{}^{\mu\nu\eta\gamma} &= J_\lambda{}^{\mu}{}_{\alpha} J_\lambda{}^{\nu}{}_{\beta} J_\lambda{}^{\eta}{}_{\rho} J_\lambda{}^{\gamma}{}_{\sigma}\hat{C}^{\alpha\beta\rho\sigma},
\end{eqnarray}
where the Jacobian matrix $J_\lambda{}^{\nu}{}_{\beta}$
transforms Euclidean coordinates to cylindrical coordinates.
The necessary and sufficient condition for $\hat{C}^{abcd}$ to be isotropic,
i.e., $\mathscr{L}_{\lambda}\hat{C}^{abcd}=0$, is:
\begin{eqnarray}
    \frac{\partial \hat{C}_\lambda{}^{\mu\nu\eta\gamma}}{\partial \lambda} = \hat{C}^{\alpha\beta\rho\sigma} \frac{\partial}{\partial \lambda}\left(J_\lambda{}^{\mu}{}_{\alpha} J_\lambda{}^{\nu}{}_{\beta} J_\lambda{}^{\eta}{}_{\rho} J_\lambda{}^{\gamma}{}_{\sigma}\right) = 0.
\end{eqnarray}
By solving this matrix equation for the constant matrix $\hat{C}^{\alpha\beta\rho\sigma}$, we get:
\begin{eqnarray}
    &\hat{C}^{1111}=\hat{C}^{3333}, \quad \hat{C}^{1133}=\hat{C}^{1122}=\hat{C}^{3322},\nonumber\\
    &\hat{C}^{1212} = \hat{C}^{3232}, \quad \hat{C}^{1221}= \hat{C}^{3333}-\hat{C}^{3232}- \hat{C}^{3322},\nonumber\\
    &\hat{C}^{1313}=\hat{C}^{3232}, \quad \hat{C}^{1331}=\hat{C}^{3333}-\hat{C}^{3232}-\hat{C}^{3322},\nonumber\\
    &\hat{C}^{2112}=\hat{C}^{3333}-\hat{C}^{3232}-\hat{C}^{3322}, \quad \hat{C}^{2121}=\hat{C}^{3232}\nonumber\\
    &\hat{C}^{2211}=\hat{C}^{2233}=\hat{C}^{3322}, \quad \hat{C}^{2222}=\hat{C}^{3333}\nonumber\\
    &\hat{C}^{2323}=\hat{C}^{3232}, \quad \hat{C}^{2332}=\hat{C}^{3333}-\hat{C}^{3232}-\hat{C}^{3322}\nonumber\\
    &\hat{C}^{3113}=\hat{C}^{3333}-\hat{C}^{3232}-\hat{C}^{3322}, \quad \hat{C}^{3131}=\hat{C}^{3232},\nonumber\\
    &\hat{C}^{3223}=\hat{C}^{3333}-\hat{C}^{3232}-\hat{C}^{3322} \quad, \hat{C}^{3311}=\hat{C}^{3322},\label{general_3dC}
\end{eqnarray}
with all other coefficients being zero. It can be seen that in 3D, the
isotropic odd elastic modulus has 3 degrees of freedom, fewer than the 6 in
2D. This is understandable as the isotropic odd elastic modulus in higher
dimensions have higher symmetry, leading to more constraints. In fact, such
high symmetry automatically imposes principal axis index symmetry on
$\hat{C}^{abcd}$: from (\ref{general_3dC}), it is evident that
$\hat{C}^{\mu\nu\eta\gamma}=\hat{C}^{\eta\gamma\mu\nu}$. Therefore, 3D odd
elastic modulus must be anisotropic. This is a consequence of fundamental
symmetry, independent of the specific physical origin of the elastic modulus.
This result is consistent with the conclusions drawn using group
representation theory in the literature~\cite{Scheibner2020}.

Similar to the 2D case, since any sufficiently small neighborhood of a point
on a 3D manifold can be considered as a 3D Euclidean space, for any 3D
elastic body reference configuration, the odd elastic modulus
$\hat{C}^{abcd}$ and $\hat{D}^{abcdef}$ are anisotropic tensor fields.
Furthermore, from a physical standpoint, elastic bodies in dimensions higher
than 3 are generally not of interest. Therefore, we assert in the main text
that due to the contribution of the anisotropic mass $\rho\Omega^{ab}$,
$\hat{C}^{abcd}$ and $\hat{D}^{abcdef}$ naturally exhibit anisotropy.
\\
\bibliographystyle{iopart-num}
\providecommand{\newblock}{}

\end{document}